%% file: 0-main.tex
\documentclass[sigconf,10pt]{acmart}
\usepackage[english]{babel}

\input{macros.tex}
\input{cmds.tex}

\renewcommand\footnotetextcopyrightpermission[1]{} % removes footnote with conference info
\setcopyright{none}

\begin{document}

\settopmatter{printacmref=false, printccs=false, printfolios=true}

\title{\sysname: Dynamic Topology Expansion for Secure and Scalable Configuration Sharing}

\author{Qianye Wang}
\affiliation{
  \institution{Peking University}
  \country{}
}

\author{Yuejie Wang}
\affiliation{
  \institution{Peking University}
  \country{}
}

\author{Yongting Chen}
\affiliation{
  \institution{Peking University}
  \country{}
}

\author{Guyue Liu}
\affiliation{
  \institution{Peking University}
  \country{}
}

% The default list of authors is too long for headers
% \renewcommand{\shortauthors}{Paper \#\paperid}

\begin{abstract}

As modern networks continue to grow in both scale and complexity, sharing real-world device configurations poses significant privacy risks—especially when adversaries can infer organizational size or resource distribution from topology data. We present NetCloak, a configuration‐anonymization framework that adaptively injects synthetic routers and hosts into the network graph to obfuscate true scale, while preserving end-to-end forwarding behavior. NetCloak’s core techniques include: (1) a graph-embedding expansion algorithm that integrates the original topology into a larger reference graph, ensuring added nodes blend seamlessly with real ones; (2) a k-degree mapping anonymity scheme that selectively adds minimal links to guarantee each original node’s degree is indistinguishable among at least k peers; (3) a mimicry-driven configuration generator that derives command templates from existing devices, preserving command ordering, naming conventions, and routing policies; and (4) a layered repair process combining SMT-based intra-AS route synthesis with iterative inter-AS filter insertion to restore protocol‐correct routing under OSPF and BGP. Extensive experiments on real and emulated campus and data-center topologies demonstrate that NetCloak effectively conceals network size—improving topological rationality by over 70\% and configuration fidelity by nearly 30\% compared to baseline methods—while reducing route-repair overhead by more than 50\% under randomized link costs. NetCloak thus enables safe, privacy-preserving configuration sharing at scale.

\end{abstract}

\maketitle
\pagestyle{plain}

\input{1-intro.tex}

\input{2-motivation.tex}
\input{3-overview.tex}

\input{4-design.tex}
\input{5-implementation.tex}
\input{6-evaluation}
\input{7-related.tex}
\input{8-conclusions.tex}

% \begin{acks}
% \end{acks}

\bibliographystyle{ACM-Reference-Format}
\bibliography{references,zotero-privnet,zotero-confmaskext}

\end{document}

%% file: macros.tex
\usepackage{hyperref}
\usepackage[compact]{titlesec}
\usepackage{enumitem}
\usepackage{multirow}
\usepackage{array}
\usepackage{graphicx}
\usepackage{caption}
\usepackage{subcaption}
\usepackage{tikz}
\usepackage{algorithm}
\usepackage[noend]{algpseudocode}
\usepackage{graphics}
\usepackage{epsfig}
\usepackage{bm}
\usepackage{booktabs}
\usepackage{enumitem}
\usepackage{multirow}
\usepackage{array}
\usepackage{graphicx}
\usepackage{caption}
\usepackage{subcaption}
\usepackage{graphics}
\usepackage{epsfig}
\usepackage{booktabs}
\usepackage{pifont}% http://ctan.org/pkg/pifont
\usepackage{xcolor, colortbl}
\usepackage{xspace}
\usepackage{listings}
\usepackage{boldline}
\usepackage{multirow}
\usepackage{xspace}
\usepackage[hang,flushmargin]{footmisc}
\usepackage[font=bf]{caption}
\usepackage{soul}
\usepackage{amsmath}
\usepackage{amsthm}
\usepackage{bbm}
\usepackage{tabularx}
\newcommand{\sysname}{NetCloak\xspace}

\definecolor{codegreen}{rgb}{0,0.6,0}
\definecolor{codegray}{rgb}{0.5,0.5,0.5}
\definecolor{codepurple}{rgb}{0.58,0,0.82}
\definecolor{backcolour}{rgb}{0.95,0.95,0.92}
\definecolor{brickred}{rgb}{0.8,0.25,0.33}
\definecolor{lightgray}{rgb}{0.83, 0.83, 0.83}

\setlist{nolistsep}
\setlist[itemize]{leftmargin=10pt}
\setlist[enumerate]{leftmargin=13pt,label={(\arabic*)}}

\lstdefinestyle{mystyle}{
    backgroundcolor=\color{backcolour},   
    commentstyle=\color{codegreen},
    keywordstyle=\color{magenta},
    numberstyle=\tiny\color{codegray},
    stringstyle=\color{codepurple},
    basicstyle=\ttfamily\footnotesize,
    breakatwhitespace=false,         
    breaklines=true,                 
    captionpos=b,                    
    keepspaces=true,                 
    numbers=left,                    
    numbersep=5pt,                  
    showspaces=false,                
    showstringspaces=false,
    showtabs=false,                  
    tabsize=2,
}
\newtheoremstyle{acmplain}%
  {.3\baselineskip\@plus.2\baselineskip
    \@minus.2\baselineskip}% space above
  {.3\baselineskip\@plus.2\baselineskip
    \@minus.2\baselineskip}% space below
  {\normalfont}% body font
  {0pt}% indent amount
  {\bfseries}% head font
  {.}% punctuation after head
  {.5em}% spacing after head
  {\thmname{#1}\thmnumber{ #2}\thmnote{ (#3)}}% head spec

\newtheoremstyle{acmdefinition}%
  {.3\baselineskip\@plus.2\baselineskip
    \@minus.2\baselineskip}% space above
  {.3\baselineskip\@plus.2\baselineskip
    \@minus.2\baselineskip}% space below
  {\normalfont}% body font
  {0pt}% indent amount
  {\bfseries}% head font
  {.}% punctuation after head
  {.5em}% spacing after head
  {\thmname{#1}\thmnumber{ #2}\thmnote{ (#3)}}% head spec

%% file: cmds.tex
\newcommand{\confmask}{ConfMask\xspace}

\newcommand{\head}[1]{\vspace{0.5mm}\noindent{\bf #1:}}
\newcommand{\sref}[1]{\S\ref{#1}}

\newcolumntype{L}[1]{>{\raggedright\let\newline\\\arraybackslash\hspace{0pt}}m{#1}}
\newcolumntype{C}[1]{>{\centering\let\newline\\\arraybackslash\hspace{0pt}}m{#1}}
\newcolumntype{R}[1]{>{\raggedleft\let\newline\\\arraybackslash\hspace{0pt}}m{#1}}

\algrenewcommand\algorithmicrequire{\textbf{Input:}}
\algrenewcommand\algorithmicensure{\textbf{Output:}}

\newcommand{\anonym}[1]{\widehat{\mbox{$#1$}\raisebox{1.8mm}{}}}

%% file: 1-intro.tex
\vspace{-1.2em}
\section{Introduction}

As networks grow in scale and complexity, the need for comprehensive configuration sharing has intensified in both operational and research contexts. In small and medium-sized enterprises, operators share configurations with external experts for performance tuning or fault diagnosis~\cite{severini-codebug-2021}. In academic settings, real-world configurations serve as foundational data for validating new algorithms and network tools~\cite{han-confighub-2024, fogel-batfish-2015, beckett-minesweeper-2017}, aiding in testing routing correctness, security policies, and failure recovery. This openness drives protocol design, diagnostic procedures, and benchmarking improvements, yet these files often reveal internal strategies, such as resource distribution, costs, and service placements, which can be exploited.

Earlier anonymization techniques~\cite{maltz-configanonym-2004, intentionet_netconan_2023, han-confighub-2024, xu_prefix-preserving_2002} focused on obfuscating direct fields like IP addresses or AS identifiers. However, modern simulation platforms~\cite{fogel-batfish-2015} and control-plane emulation frameworks~\cite{neumann_book_2015} can reconstruct topology and routing properties, exposing host relationships or traffic paths. These challenges demand deeper anonymization strategies that guard indirectly inferable details.

\confmask~\cite{wang-confmask-2024} addresses inferred topology and route privacy by inserting fake links and hosts within a functional equivalence model. However, it cannot introduce additional router nodes, leaving organizations that wish to hide actual network size underserved. For example, data centers seeking to mask true capacity might find IP anonymization and link manipulation insufficient, while research groups without access to high-scale configurations cannot readily scale up partial datasets.

We present an anonymization scheme that adaptively inserts new network nodes to obscure scale-related insights. This extension of existing approaches faces significant challenges:

\begin{itemize}
    \item \head{Rational Topology Generation}  
    Unusual structures can expose added nodes or edges, enabling adversaries to identify real routers. Expansions must conform to normal graph properties. While fake nodes offer inherent anonymity, standard $k$-anonymity~\cite{liu-kDA-2008} does not exploit this, and applying it naively may cause excessive edge modifications that degrade realism, especially in networks with varying node degrees.

    \item \head{Generating Fake Node Configurations}  
    Graph anonymization research~\cite{liu-kDA-2008, zhou-kNA-2008, zou-kauto-2009, jorgensen-dp-edge-2016} typically targets social networks with minimal attributes. In contrast, routers carry complex settings---routing protocols, hardware constraints, QoS policies---that must seamlessly integrate with existing configurations. Merely adding nodes is insufficient; IP addresses, interfaces, and routing protocols must be consistently assigned. Discrepancies in command style or naming can expose inserted nodes.

    \item \head{Efficient Iterative Repairs}  
    \confmask iteratively compares forwarding tables, adding filters or adjusting link costs until route discrepancies vanish. In large expansions with many fake devices, each iteration becomes costly. SMT-based synthesis~\cite{el-hassany-netcomplete-2018, jacobson-cpr-2017} can encode routing constraints but often struggles with advanced cross-protocol scenarios. Balancing iterative adjustments with SMT-based solutions is crucial for stable performance and accurate route restoration.
\end{itemize}

We introduce \sysname, a configuration anonymization system that extends network scale using a graph embedding approach while preserving functional equivalence. Its key contributions include:

\begin{itemize}
    \item \head{Topology Expansion and Anonymization (\sref{sec:design-1}, \sref{sec:design-2})}  
    We propose a graph-embedding-based expansion method and a new $k$-anonymity definition, leveraging the anonymity gained from added routers while minimizing edge modifications to maintain a coherent topology (\sref{sec:overview-definitions}).

    \item \head{Mimicry-Based Configuration Generation (\sref{sec:design-3})}  
    We design a mimicry system that generates configuration files for new routers by reusing real router configurations as templates, preserving command ordering, stanzas, and naming schemes. Filter adaptations also align fake hosts' traffic patterns with those of real hosts, ensuring style consistency and functional fidelity.

    \item \head{Layered Routing Repair Framework (\sref{sec:design-4})}  
    Within each AS, SMT-based constraints repair routes while balancing performance under varied link costs. Across AS boundaries, iterative comparisons of border-router forwarding tables and filter additions offer flexible handling of cross-AS route advertisements or BGP decisions.
\end{itemize}

We evaluate \sysname on real and emulated networks. Compared to a baseline, it improves topological rationality by 73.5\%, increases configuration similarity by 29.0\%, and accelerates routing repairs by 60.0\% under random link costs, allowing broader configuration sharing without revealing private scale attributes. Moreover, \sysname preserves functional correctness across protocols, ensuring that real traffic behaviors remain intact after anonymization.

%% file: 2-motivation.tex
%
% 2 Motivation
%
% 2.1 Motivation Scenarios
%   * Collaborative debugging
%   * Sharing configuration for research purposes
% 2.2 Sensitive Configuration Information
%   * Network topology
%   * Routing paths
%   * Personally identifiable information
%   * Limitations of existing approaches
%
\section{Background}
\label{sec:motivation}

In this section, we highlight scenarios to demonstrate the need for scalable network configuration sharing (\sref{sec:motivating-scenarios}). We then discuss limitations of existing anonymization approaches (\sref{sec:motivation-limitations}) and sensitive information in configurations (\sref{sec:motivation-threat-model}).

\subsection{Motivating Scenarios}\label{sec:motivating-scenarios}

\head{Collaborative Debugging}
Small to medium-sized organizations often share partial network configurations with third parties to diagnose issues~\cite{wang-confmask-2024,noauthor_cisco_2011,garros_state_2020}. Sharing full configurations risks exposing sensitive business data such as passwords, topology details, and overall network scale. Existing methods focus on removing PII like passwords, IP addresses, and AS numbers~\cite{slagell_network_2004,xu_prefix-preserving_2002,minshall_tcpdpriv_1997,intentionet_netconan_2023} or anonymizing routing paths~\cite{wang-confmask-2024}, yet they retain the original router count. This can reveal an organization’s true scale and resource allocation, underscoring the need for an anonymization solution that also conceals network scale.

\head{Research Datasets}
There is growing research on network configurations for verification~\cite{beckett-minesweeper-2017,khurshid_veriflow_2012,zhang_differential_2022}, synthesis~\cite{el-hassany-netcomplete-2018,el-hassany-syNET-2017,ramanathan_practical_2023,beckett_network_2017}, and repair~\cite{jacobson-cpr-2017,yang_diagnosing_2023,liu_automatic_2024}. However, most evaluation datasets are synthesized, while real-world configurations are scarce due to privacy concerns. Real configurations often involve only a limited number of nodes (e.g., 10 for Internet2~\cite{noauthor_internet2_nodate}), constrained by production environments~\cite{han-confighub-2024} and the effort required to remove sensitive data. Therefore, methods to scale up both configuration style diversity and node count are urgently needed.

\subsection{Limitations of Existing Approaches}\label{sec:motivation-limitations}

Previous configuration anonymization work generally falls into two categories:  
\textbf{(i)} Some approaches focus on anonymizing \textbf{sensitive configuration fields} that may leak business or personal data~\cite{intentionet_netconan_2023,xu_prefix-preserving_2002,han-confighub-2024}. For example, public IP addresses are anonymized using prefix-preserving techniques~\cite{xu_prefix-preserving_2002}, ensuring that addresses sharing the same k-bit prefix before anonymization retain that prefix afterward, thus preserving the IP space’s structure. Similarly, AS numbers are anonymized with adjustments to regular expressions~\cite{intentionet_netconan_2023,han-confighub-2024}, and fields such as administrative login information, route-map names, and descriptions are replaced with hashed values~\cite{intentionet_netconan_2023}. However, these tools consider only data directly extractable from configuration files.

\textbf{(ii)} Building on this, \confmask~\cite{wang-confmask-2024} targets \textbf{indirectly disclosed information} (e.g., network topology and traffic paths inferred via Batfish) that may reveal traffic patterns and node relationships.

For example, in a campus network, \confmask extracts the topology (Fig.~\ref{fig:motivation-campus-origin}) and anonymizes it by adding fake links (red lines in Fig.~\ref{fig:motivation-campus-confmasked}). These fake links disrupt the original routing paths, necessitating repair. In the original network, router \texttt{r5} uses \texttt{r4} as the next-hop for destination \texttt{h1}. After anonymization, a shorter path leads \texttt{r5} to select \texttt{r1}. To correct this, \confmask increases the cost of fake link \texttt{A} (or adds BGP filters) so that \texttt{r5} returns to \texttt{r4}. However, this may force \texttt{r5} to choose \texttt{r2}, requiring further cost adjustments on fake link \texttt{C}. This iterative process repeats until the original routing paths are restored.

The anonymization approach of \confmask faces several issues:

\begin{figure}[!t]
     \centering
     \begin{subfigure}[t]{0.32\columnwidth}
        \centering
        \includegraphics[width=.85\linewidth]{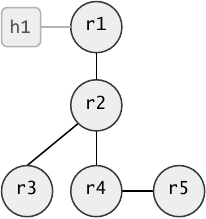}
        \caption{}
        \label{fig:motivation-campus-origin}
     \end{subfigure}
     \hfill
     \centering
     \begin{subfigure}[t]{0.32\columnwidth}
        \centering
        \includegraphics[width=.85\linewidth]{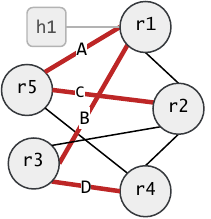}
        \caption{}
        \label{fig:motivation-campus-confmasked}
     \end{subfigure}
     \hfill
     \centering
     \begin{subfigure}[t]{0.32\columnwidth}
        \centering
        \includegraphics[width=.7\linewidth]{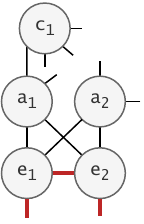}
        \caption{}
        \label{fig:motivation-fattree}
     \end{subfigure}

    \vspace{-1em}

     \caption{Two example networks. (a) Sample Campus Network; (b) Anonymized Campus Network with $k=3$ Topology Anonymity; (c) Sub-topology of FatTree-04 network. Red edges are the fake links added by \confmask.}
\end{figure}

\head{Issue 1: Inability to Anonymize Network Scale}  
\confmask only modifies existing configuration files by adding fake links; it cannot add new nodes or generate complete configurations. A naive solution is to combine \confmask with configuration generation tools (e.g., NetComplete). However, such tools rebuild network configurations from scratch based on intended policies and do not support networks with preexisting configurations. In particular, NetComplete supports only a limited set of commands (e.g., OSPF cost, route policy permit/deny) and omits many functional commands.

\begin{lstlisting}[caption={Config with \texttt{peer-group} Characteristics}, label={lst:motivation-issue1-original}]
router bgp 100
  neighbor ISP1 peer-group
  neighbor ISP1 remote-as 714
  neighbor ISP1 route-map ISP-IMP in
  neighbor 17.56.7.8 peer-group ISP1
  neighbor 17.56.8.9 peer-group ISP1
\end{lstlisting}
\vspace{-1em}
\begin{lstlisting}[caption={NetComplete Synthesized Configuration}, label={lst:motivation-issue1-netcomplete}]
router bgp 100
  neighbor 17.56.7.8 remote-as 714
  neighbor 17.56.7.8 route-map imp-p1 in
  neighbor 17.56.8.9 remote-as 714
  neighbor 17.56.8.9 route-map imp-p2 in
\end{lstlisting}
\vspace{-.8em}

For example, the \texttt{peer-group} command in an existing edge router configuration (Listing~\ref{lst:motivation-issue1-original}) reflects bulk management, but the configuration synthesized by NetComplete does not support \texttt{peer-group} and treats each neighbor individually (Listing~\ref{lst:motivation-issue1-netcomplete}). Omitting commands such as \texttt{next-hop-self} and \texttt{redistribute} can change routing behavior. Moreover, these tools produce configuration files based on predefined templates, making them stylistically distinct from the originals and prone to deanonymization.

\head{Issue 2: Potentially Ineffective Topology Anonymity or Irrational Topologies}  
\confmask employs an edge-adding $k$-degree topology anonymization method. In networks with certain node degree distributions, this approach offers limited anonymity. For instance, in a FatTree topology with a highly regular degree distribution (Fig.~\ref{fig:motivation-fattree}), if the anonymity goal $k$ is small, \confmask does not modify the upper-layer nodes and barely alters the lower layers, yielding minimal anonymity. Conversely, in a power-law distributed campus network (Fig.~\ref{fig:motivation-campus-origin}), where few nodes have very high degrees, \confmask adds many fake links to achieve $k$-degree anonymity (Fig.~\ref{fig:motivation-campus-confmasked}), disrupting structural regularity and resulting in an irrational topology (as defined in~\sref{sec:overview-definitions}).

\head{Issue 3: Potential Efficiency Problems with Iterative Repair}  
To repair routing paths, \confmask assigns costs (or adds BGP route policies) to fake links by referencing original path costs between real routers. This method is unsuitable for links involving new routers, as the cost for new paths is unknown. A naive approach sets the cost high enough so that the fake link is excluded from routing tables; however, this pattern is easily detected and vulnerable to deanonymization. \confmask instead iteratively increases fake link costs and verifies their exclusion, but this forward-table-based repair process may require an uncontrollable number of iterations to converge.

\subsection{Threat Model and Sensitive Information}\label{sec:motivation-threat-model}

Network owners hesitate to disclose original configurations due to privacy risks, legal concerns, or policy violations. When a network owner (A) shares configurations with a third party (B) for troubleshooting or research, B may act as an adversary by (1) accessing shared configurations, (2) inferring network behavior, and (3) leveraging network analysis tools (e.g., simulations)~\cite{wang-confmask-2024}. In this work, we focus on two types of sensitive information:

\begin{itemize}
    \item \head{Network Scale} The network scale can reveal an organization’s size, resource allocation, and operational status. Even if fake links obscure detailed topology, an adversary can combine network scale data with external sources (e.g., organizational charts, employee counts, or business distribution) to infer part or all of the true topology. This correlation risk makes network scale particularly sensitive.
    \item \head{Configuration Style} Template-based synthesis tools generate configurations for new nodes that differ stylistically from original files. Adversaries can exploit these unique stylistic patterns to identify and remove fake nodes, exposing the genuine configuration for further attacks. Maintaining consistency in configuration style is therefore crucial.
\end{itemize}

Protecting both network scale and configuration style is essential, as leakage in either area can enable attackers to reconstruct or de-anonymize the network, posing significant security risks.

%% file: 3-overview.tex
\section{System Overview}
\label{sec:overview}
\begin{figure*}[ht]
    \centering
    \includegraphics[width=1.0\textwidth]{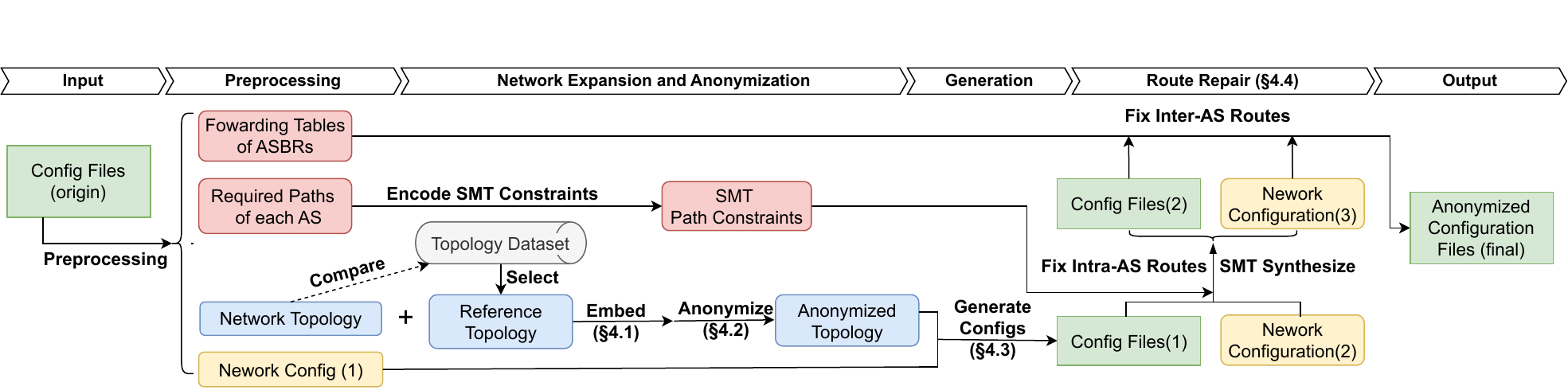}
    \vspace{-2em}
    \caption{\sysname Workflow}
    \vspace{-1em}
    \label{fig:workflow}
\end{figure*}

We present \textbf{\sysname}, a configuration anonymization system that expands the network while preserving functional equivalence. Key design ideas are:

\head{Idea 1: Node Addition via Graph Embedding}  
Our approach selects a real-world reference topology larger than the original network, extracts its degree sequence, and generates an expanded topology containing the original as a subgraph. By closely matching the reference’s degree sequence, the expanded network retains high rationality.

\head{Idea 2: k-Degree Mapping Anonymity (k-DMA)}  
Building on graph embedding, k-DMA targets a threat model where attackers know only the original network’s degree distribution. Since embedding alters node degrees, k-DMA ensures each original node’s mapping in the anonymized network is at least k-indistinguishable. Unlike traditional k-degree anonymity, k-DMA makes fewer topological changes, preserving higher rationality.

\head{Idea 3: Mimicry-Based Configuration Generation}  
We simulate network growth to convert new nodes and edges into router and link configurations. For each fake router, a real router with similar configuration traits is chosen as a template, and critical fields are replaced or filled in. For fake hosts, a mimicry-driven routing policy treats their traffic the same as that of real hosts.

\head{Idea 4: Network Repair Using SMT and Iterative Methods}  
Repair is divided into intra-AS and inter-AS stages. Within each AS, an SMT-based approach encodes original path constraints, reducing iterations compared to pure iterative methods. For inter-AS repair, we compare anonymized and original forwarding tables of AS boundary routers and add filters as needed, avoiding the complexity of modeling BGP and route redistribution in SMT.

% \vspace{-.8em}

\subsection{Definitions}\label{sec:overview-definitions}
\vspace{-.2em}
We adopt definitions from \confmask~\cite{wang-confmask-2024}, representing a network configuration as
\(
\mathrm{CFG} = (G, DP).
\)
Here, \(G = (V, E)\) is the network-layer topology, where each router is a node in \(V\) and each Layer\,3 link is an edge in \(E\). The data plane \(DP\) contains all host-to-host routing paths (e.g., via traceroute), with each path \(p = (h_s, r_s, \dots, r_d, h_d)\) including the source and destination hosts, their egress routers, and any intermediate routers. The anonymized configuration, topology, and data plane are denoted as \(\anonym{CFG}\), \(\anonym{G}\), and \(\anonym{DP}\), respectively.

We now present our k-anonymity definitions. 

\begin{definition}\label{def:weak-k-deg-anonymity}
\textbf{(Weak k-Degree Mapping Anonymity)}\\  
For any node \(v \in V\) in \(G\) with degree \(\deg(v)\), if there exist at least \(k\) nodes in \(\anonym{G}\) whose degrees are at least \(\deg(v)\), then \(v\) is \emph{weakly k-degree mapping anonymous}. If this holds for every node, then \(\anonym{G}\) satisfies weak k-degree mapping anonymity.
\end{definition}

\begin{definition}\label{def:strong-k-deg-anonymity}
\textbf{(Strong k-Degree Mapping Anonymity)}\\  
Order the nodes in \(G\) in descending order as \(v_1, v_2, \dots, v_n\) with \(\deg(v_1) \ge \deg(v_2) \ge \dots \ge \deg(v_n)\). For the \(i\)-th node \(v_i\), if there exist at least \((k+i-1)\) nodes in \(\anonym{G}\) with degrees at least \(\deg(v_i)\), then \(v_i\) is \emph{strongly k-degree mapping anonymous}. If this holds for all nodes, \(\anonym{G}\) satisfies strong k-degree mapping anonymity.
\end{definition}

In general, \emph{k-degree mapping anonymity} applies when the original topology is preserved and expanded, ensuring that original node degrees do not decrease. Unlike conventional k-degree anonymity—which assumes attackers observe the anonymized degree sequence—k-degree mapping anonymity assumes attackers know only the original degree sequence. Weak k-degree mapping anonymity requires at least \(k\) plausible mappings per node, whereas strong k-degree mapping anonymity guarantees that even if some nodes are identified, each remaining node has at least \(k\) indistinguishable possibilities based solely on degree. This paper focuses on strong k-degree mapping anonymity, as it limits the attacker's advantage by relying solely on the original network's degree information.

Finally, we define \emph{topology rationality} based on degree sequence.

\begin{definition}\label{def:topology-rationality}
\textbf{(Topology Rationality)}
Let \(G_{\mathrm{ref}}\) be the reference topology chosen for anonymization. The \emph{degree-sequence-based topology rationality} is defined as the Kolmogorov--Smirnov (K-S) distance between the degree sequences of \(\anonym{G}\) and \(G_{\mathrm{ref}}\).
\end{definition}

\vspace{-.4em}

\subsection{Workflow}

The workflow of \sysname comprises four phases: Preprocessing, Network Expansion \& Anonymization, Configuration Generation, and Route Repair (see Fig.~\ref{fig:workflow}). \sysname strives to achieve the same routing utility as \confmask by preserving network functional equivalence~\cite{wang-confmask-2024}. Functional equivalence guarantees that each original data-plane path is maintained via an injective mapping in the anonymized configuration, while strong functional equivalence (SFE) further enforces protocol-specific rules for distance-vector, BGP, and link-state protocols. In essence, SFE ensures that every original route remains valid and that fake links do not alter observed routing decisions, thus implying functional equivalence~\cite{wang-confmask-2024}.

\noindent \textbf{Preprocessing.}  
\sysname begins by parsing the uploaded configuration files to reconstruct the network topology and extract pertinent configuration details. It then computes and stores the data necessary for route repair, including the routing path requirements for each AS and the forwarding tables of all ASBRs.

\noindent \textbf{Network Expansion \& Anonymization.}  
Based on the user-specified number of routers to add, \sysname selects an appropriate \emph{reference topology} from a real-world network. It performs node mapping, computes the expected degree sequence, and constructs an embedded graph that contains the original topology as a subgraph to guarantee functional equivalence. Finally, a \emph{k-anonymization} algorithm is applied to the embedded graph to yield the anonymized topology.

\noindent \textbf{Configuration Generation.}  
In this phase, the topological updates are translated into router and link configuration changes. For each newly added fake router, \sysname selects a template from a corresponding real router and updates that template with synthesized configuration data. It then iterates through all router configuration files to adjust filter rules based on the mapping between fake and real hosts, ensuring that traffic from fake hosts is handled identically to that from real hosts.

\noindent \textbf{Route Repair.}  
Within each AS, an SMT-based method is employed to enforce the precomputed route requirements by encoding path constraints into an SMT solver, which derives and applies the necessary configuration changes. For inter-AS routing, \sysname iteratively compares each ASBR’s current forwarding table with the stored version and resolves discrepancies by adding filters. This process enforces functional equivalence within each AS and across AS boundaries, ensuring compliance with BGP SFE conditions.

%% file: 4-design.tex
% \vspace{-3em}
\section{Key Design}
\label{sec:design}

This section introduces the core design of \sysname, encompassing network expansion, network anonymization, a mimicry-based configuration generation system, and SMT and iterative based network repair framework.

\subsection{Network Expansion}
\label{sec:design-1}

\begin{figure}[h!]
    \centering
    \includegraphics[width=0.45\textwidth]{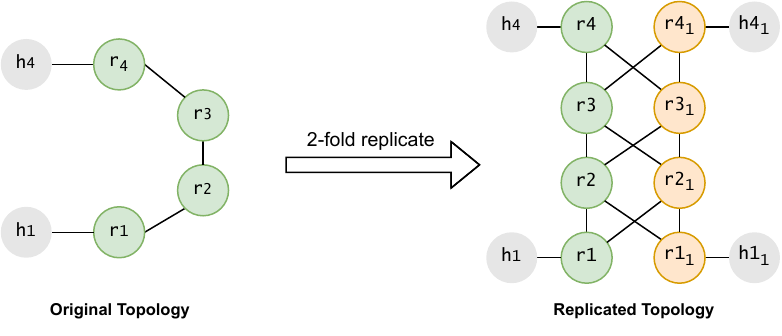}
    \vspace{-1em}
    \caption{Replica Method}
    \label{fig:replica}
\end{figure}

\head{Strawman 1: Replica}
A straightforward way to expand a network is to replicate the entire topology. Replication-based algorithms can actually provide strong anonymity under certain threat models. For instance, Takbiri et al. proposed a $k$-fold replication algorithm~\cite{takbiri-replica-2019} that yields a $k$-automorphic anonymized graph. In each iteration, this algorithm creates a fake node for every real node, mirroring the same neighbors, and repeats for \(k-1\) iterations until the $k$-automorphic property is satisfied~\cite{zou-kauto-2009}. 

Fig.~\ref{fig:replica} shows the result of 2-fold replication on a small network of four routers (r1--r4) and two hosts (h1, h4). All fake nodes (\(ri_1\)) share the same neighbors as their corresponding real nodes (\(ri\)), producing a 2-automorphic graph with automorphism \(f(\mathrm{ri}) = \mathrm{ri}_1\).

This $k$-fold replication strategy resists \emph{seed-based attacks}, where an adversary knows \(\sigma(v)\) for some subset \(V_s \subset V\) and aims to identify \(\sigma(v)\) for nodes in \(V - V_s\)~\cite{takbiri-replica-2019}. In the worst case, if mappings for \(n-1\) nodes are known, the final node still has \(k\) equally likely candidates, forcing a \(1/k\) guess probability. This is a stronger threat model than the degree-based de-anonymization considered here. While $k$-fold replication alone can ensure strong anonymity, it has several drawbacks:

\noindent(1) \textbf{Unrealistic topologies.} Replicating nodes and edges creates a network with obvious structural patterns. Such replicas make it easier for adversaries to detect the anonymization scheme and launch targeted de-anonymization.

\noindent(2) \textbf{Limited flexibility.} The method can only add nodes in multiples of the original size, preventing fine-grained control over how many nodes to add.

\noindent(3) \textbf{Weaker structural protection.} Once attackers recognize $k$-fold replication, they can group nodes by adjacency to strip away the added replicas. Our own tests confirm that unless the original topology is inherently automorphic, an attacker can reconstruct an unlabeled version of the real topology from the anonymized graph alone, without extra knowledge. The anonymity guarantee degrades to a state where each unlabeled node corresponds to a group of \(k\) indistinguishable nodes in the anonymized network.

Overall, $k$-fold replication can theoretically provide $k$-automorphic anonymity guarantees, but its topological artifacts and rigidity in node addition diminish practical usefulness. As such, we seek more flexible approaches that yield realistic topologies and finer anonymity controls.

\begin{figure*}[ht]
    \centering
    \includegraphics[width=.95\textwidth]{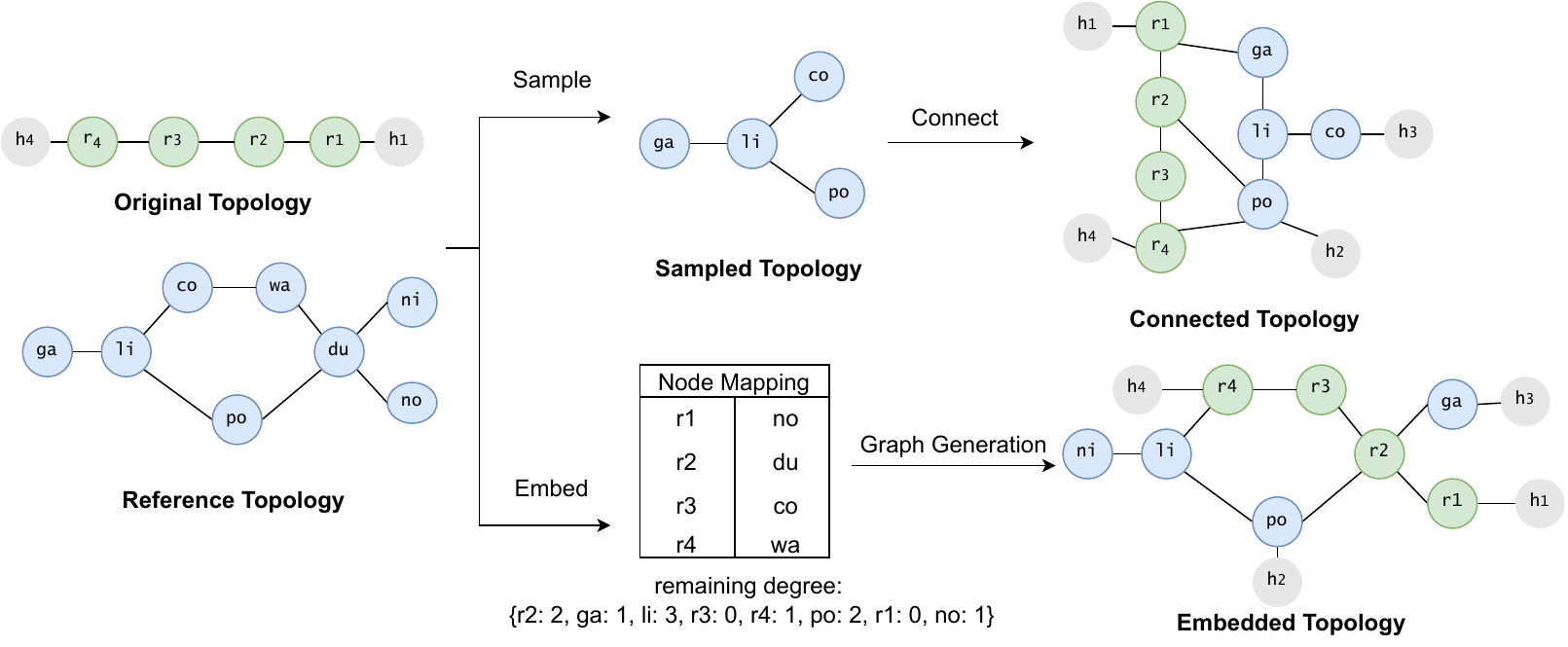}
    \vspace{-1em}
    \caption{Sample-Connect and Embedding Method}
    \vspace{-.8em}
    \label{fig:sample-embed}
\end{figure*}

\head{Strawman 2: Sample-Connect}
To obtain a more realistic topology than that produced by the replica method, our second strawman approach selects an existing real-world network topology as a reference. We then employ graph sampling to extract a subgraph containing the desired number of nodes from this reference, and finally connect the original network with the sampled subgraph via randomly added edges. Compared to Strawman 1, the resulting topology exhibits weaker anonymization signatures and higher topological rationality. Fig.~\ref{fig:sample-embed} illustrates the sample-connect workflow, assuming four extra nodes are added using a reference topology of eight nodes.

A critical step in Strawman 2 is choosing an appropriate sampling algorithm. Mainstream techniques generally fall into two categories:
\begin{itemize}
    \item \textbf{Traversal-based} sampling algorithms such as Breadth-/Depth-/Random- First Sampling (B-/D-/R-FS)~\cite{doerr-bdrfs-2013}, Snow-Ball Sampling (SBS)~\cite{leo-sbs-1961}, and Forest Fire Sampling (FFS)~\cite{leskovec-ff-2005,leskovec-ffs-2006}.
    \item \textbf{Random-Walk-based} sampling algorithms, including Metropolis Hastings Random Walk (MHRW)~\cite{metropolis-mh-1953,upfall-mhrw-2005}, Metropolis Hastings Delayed Acceptance Sampling (MHDA)~\cite{lee-MHDA-2012}, and Rejection Controlled Metropolis-Hastings Random Walk (RCMH)~\cite{li-rcmh-2015}.
\end{itemize}

Most existing techniques focus on approximating the original graph's properties rather than generating subgraphs that closely resemble the original. Therefore, optimizations tailored for property estimation may not be effective for our subgraph-generation task. To evaluate this, we compared the degree-sequence similarity between sampled subgraphs and the original graph, measured by K-S distance. Experiments were conducted on TopoZoo datasets, with each sampling method tested 1,000 times; results are presented in Table~\ref{tab:sample-traversal} and Table~\ref{tab:sample-rw}.

\begin{table}[h!]
\centering
\begin{subtable}[h]{0.48\textwidth}
\centering
\begin{tabular}{|c|c|c|c|c|c|}
\hline
Sampling Rate & BFS       & DFS       & RFS       & SBS       & FFS       \\ \hline
0.25          & 0.3197    & 0.3163    & 0.3199    & 0.2873    & 0.2866    \\ \hline
0.5           & 0.2691    & 0.2624    & 0.2668    & 0.2487    & 0.2423    \\ \hline
0.75          & 0.2380    & 0.2323    & 0.2370    & 0.2397    & 0.2424    \\ \hline
\end{tabular}
\caption{Traversal-Based Sampling Method.}
\label{tab:sample-traversal}
\end{subtable}
\hfill
\begin{subtable}[h]{0.48\textwidth}
\centering
\begin{tabular}{|c|c|c|c|c|}
\hline
Sampling Rate & RW        & MHRW      & MHDA      & RCMH      \\ \hline
0.25          & 0.3224    & 0.3192    & 0.3226    & 0.3209    \\ \hline
0.5           & 0.2006    & 0.1981    & 0.2049    & 0.1993    \\ \hline
0.75          & 0.1393    & 0.1377    & 0.1415    & 0.1379    \\ \hline
\end{tabular}
\caption{Random-Walk Based Sampling Method}
\label{tab:sample-rw}
\end{subtable}
\caption{Comparison of sampling methods and sampling rates.}
\vspace{-1.5em}
\label{tab:combined-sampling}
\end{table}

Our results indicate that a higher sampling ratio generally yields a subgraph that better aligns with the original graph’s properties. Under high sampling ratios, random-walk-based algorithms clearly outperform traversal-based methods. Moreover, specialized optimizations for random-walk-based sampling, originally designed for property estimation, did not significantly improve performance over standard Random Walk sampling in our scenario. Consequently, our reference topology selection algorithm prioritizes topologies whose sizes are close to the desired number of nodes, ensuring the sampling algorithm operates at a high ratio. Unless specified otherwise, we use standard Random Walk sampling by default.

Nevertheless, the random connect step in sample-connect may still result in lower topological rationality. To address this, we propose a further refined solution.

\head{Our Approach: Embedding}
To improve topology rationality, our embedding method establishes a node mapping between the original topology and a reference topology based on degree relationships, then uses the residual degree sequence to complete the embedded graph. Since topology rationality is integrated into the construction objective, this approach is expected to outperform sample-connect. The method comprises two steps: Node Mapping and Graph Embedding.

\head{(i) Node Mapping} For each node \(v_{\mathrm{ori}}\) in the original network, we select a node \(v_{\mathrm{ref}}\) in the reference topology satisfying \(\deg_{G_{\mathrm{ref}}}(v_{\mathrm{ref}}) \ge \deg_{G_{\mathrm{ori}}}(v_{\mathrm{ori}})\). The expected degree of \(v_{\mathrm{ori}}\) in the anonymized graph is then set to \(\deg_{G_{\mathrm{ref}}}(v_{\mathrm{ref}})\). This selection is formulated as a complete matching problem in a bipartite graph \(G=(U,V,E)\), where:
\begin{itemize}
    \item \(U\) is the set of nodes in \(G_{\mathrm{ori}}\).
    \item \(V\) is the set of nodes in \(G_{\mathrm{ref}}\).
    \item \(E = \{(u, v) \in U \times V \mid \deg_{G_{\mathrm{ori}}}(u) \leq \deg_{G_{\mathrm{ref}}}(v)\}\), where \(\deg_{G_{\mathrm{ori}}}(u)\) denotes the degree of \(u\) in \(G_{\mathrm{ori}}\) and \(\deg_{G_{\mathrm{ref}}}(v)\) denotes the degree of \(v\) in \(G_{\mathrm{ref}}\).
\end{itemize}

A maximum matching algorithm (e.g., the Hungarian algorithm) is employed to solve it. If the maximum matching size is less than \(|U|\), the mapping fails and a different reference topology must be chosen.

\begin{algorithm}[!ht]
\caption{Greedy Graph Embedding Algorithms}
\label{alg:embed-greedy}
\begin{algorithmic}[1]

\Require $G$: original graph, $G_{ref}$: reference graph
\Ensure $G_{emb}$: embedded graph

\vspace{4pt}
\Function{edge\_completion}{}
\State \textbf{while} True \textbf{do}
\State \quad pick node $u$ with the largest $need[u]$
\State \quad \textbf{for} $v$ in other nodes (sorted by $need[v]$) \textbf{do}
\State \qquad \textbf{if} add edge($u,v$) is possible \textbf{then}
\State \qquad\quad add edge($u,v$); $need[u]{-}{-},\, need[v]{-}{-}$
\State \quad \textbf{if} no edge is added \textbf{then return}
\EndFunction

\vspace{4pt}
\Function{edge\_rearrangement}{}
\State \textbf{define} $diff(u) = targetDeg[u] - degree(u)$
\State \textbf{while} True \textbf{do}
\State \quad $under\_nodes \leftarrow \{\,u \mid diff(u) > 0\}$
\State \quad \textbf{for each} pair $(u,v)$ in $under\_nodes$ \textbf{do}
\State \qquad \textbf{if} find a possible edge $(x,y)$ \textbf{then}
\State \qquad\quad remove $(x,y)$; add $(u,x)$ \& $(v,y)$
\State \quad \textbf{if} no swap is found \textbf{then return}
\EndFunction

\vspace{4pt}
\Function{embed\_graph\_greedy}{}
\State $match \leftarrow \text{complete\_match}(G\_degs, ref\_degs)$
\State \textbf{if} $match$ is $None$ \textbf{then return} $None$
\State create $G_{emb}$ with all all edges of $G$
\State define $targetDeg[u]$ from degrees in $G_{ref}$
\State compute $need[u] = targetDeg[u] - deg(u)$ in $G_{emb}$
\State call \Call{edge\_completion}{}
\State call \Call{edge\_rearrangement}{}
\State \textbf{return} $G_{emb}$
\EndFunction

\end{algorithmic}
\end{algorithm}

\head{(ii) Graph Embedding} As shown in Algorithm~\ref{alg:embed-greedy}, the embedding process first adds all nodes from \(G_{\mathrm{ref}}\) and edges from \(G_{\mathrm{ori}}\) into \(G_{\mathrm{anonym}}\) according to the established mapping, then computes the remaining expected degrees. Since the resulting degree sequence may not be directly realizable as a simple graph—and the process must extend an existing subgraph—traditional graph generation models are not applicable. Instead, we propose an edge-addition procedure inspired by the Havel–Hakimi algorithm that iteratively connects the node with the highest remaining degree until no more edges can be added, as shown in Algorithm~\ref{alg:embed-greedy}.

To further align the degree sequence of \(G_{\mathrm{anonym}}\) with \(G_{\mathrm{ref}}\), a greedy edge rearrangement is performed. For any two nodes \(u, v\) that have not reached their expected degrees, let \(S_u\) and \(S_v\) be the sets of nodes not adjacent to \(u\) and \(v\), respectively. If nodes \(x\in S_u\) and \(y\in S_v\) are connected by an edge \((x,y)\) in \(G_{\mathrm{anonym}}\), we remove \((x,y)\) and add \((u,x)\) and \((v,y)\), reducing the degree gap for both \(u\) and \(v\) without affecting others. This rearrangement repeats until no further rewiring is possible.

Fig.~\ref{fig:sample-embed} illustrates a topology produced by graph embedding. In the example, after node mapping and incorporating original edges, the remaining degrees are \(\{\text{r2}:2, \text{ga}:1, \text{li}:3, \text{r3}:0, \text{r4}:1, \text{po}:2, \text{r1}:0, \text{no}:1\}\). Since this sequence is realizable, our algorithm outputs an anonymized graph whose degree sequence exactly matches that of the reference topology.

% \vspace{-.8em}

\subsection{Network Anonymization}
\label{sec:design-2}

The graphs constructed via either Sample-Connect or Embedding must be further anonymized to defend against attacks based on degree knowledge. If we assume a threat model in which the attacker can learn each node’s degree after anonymization, then we still need to apply k-degree anonymity. However, if we assume a weaker threat model in which the attacker’s knowledge of node degrees pertains only to the original graph, then the embedding process already provides some extent of anonymity for original nodes by potentially increasing their degrees. In such scenarios, we only need to perform k-degree mapping anonymity (k-DMA) to achieve effective k-anonymity while preserving higher topological rationality. We implement a greedy k-DMA anonymization algorithm as shown in Algorithm~\ref{alg:kdma-greedy}.

\begin{algorithm}[!ht]
\caption{Greedy $k$-Degree Mapping Anonymize}\label{alg:kdma-greedy}
\begin{algorithmic}[1]

\Require Graph $G$, embedded graph $G_{\text{emb}}$, parameter $k$, anonymity level $kdma\_level$
\Ensure $G_{\text{anonym}}$ satisfying $k$-DMA

\vspace{4pt}
\State $G_{\text{anonym}} \leftarrow$ copy of $G_{\text{emb}}$
\State\textbf{for} {each degree $d_i$ in $sorted\_degrees$}
\State \quad compute $needed\_number$ based on $kdma\_level$ 
\State \quad $candidates \leftarrow$ sort all nodes $u$ in $G_{\text{anonym}}$ with $\mathrm{deg}(u) < d_i$
\State\textbf{for} {each $cand\_node$ in $candidates$}
\State \quad\quad \textbf{if} $needed\_number \le 0$ \textbf{then break}
\State \quad\quad $shortfall \leftarrow d_i - \mathrm{deg}(cand\_node)$
\State \quad\quad pick connection targets for $cand\_node$
\State \quad\quad add edges until $cand\_node$'s degree reaches $d_i$ 
\State \Return $G_{\text{anonym}}$

\end{algorithmic}
\end{algorithm}

This algorithm processes the degree sequence of \(G_{\mathrm{ori}}\) in descending order. For each degree \(d_i\), it computes how many nodes \(n_{\mathrm{need}}\) in \(G_{\mathrm{anonym}}\) must be elevated to \(d_i\) based on the current anonymization level and the current degree sequence of \(G_{\mathrm{anonym}}\). It then traverses \(G_{\mathrm{anonym}}\) in descending degree order and, for candidate nodes \(v_{\mathrm{cand}}\) whose degree is below \(d_i\), randomly adds edges until \(v_{\mathrm{cand}}\) reaches degree \(d_i\). This is repeated for \(n_{\mathrm{need}}\) nodes.

If maximizing topological rationality is critical, we also offer an all-in-one MaxSMT-based approach that simultaneously handles graph embedding and k-degree mapping anonymity. Let \(\text{match}[v]\) be the node in \(G_{\mathrm{ref}}\) matched to \(v\) in \(G_{\mathrm{ori}}\), and let \(\mathrm{x}_{r_1, r_2}\) indicate whether an edge exists between \((r_1, r_2)\) in \(G_{\mathrm{anonym}}\). We encode the following \textit{hard constraints} in the MaxSMT problem:

    \noindent (1) Preserve original edges:
    \[
        \forall\, (u, v) \in G_{\mathrm{ori}}: \quad \mathrm{x}_{(\text{match}(u), \text{match}(v))} = \mathrm{True}.
    \]
    
    \noindent (2) Define node degree: 
    \(
        \forall\, r:\quad \mathrm{degExpr}(r) = \sum_{r' \neq r} \mathrm{x}_{(r, r')}.
    \)
    
    \noindent (3) Satisfy $k$-DMA: Sort the nodes of $G_{\text{ori}}$ in descending order of degree, $\{v_1, v_2, \dots, v_n\}$. Then:
    \[
        \forall\, i \in \{1, \dots, n\}: \quad \sum_{r} \left[ \mathrm{degExpr}(r) \ge \deg_G(v_i) \right] \ge (k + i - 1).
    \]

To minimize the difference between the degree sequence of $G_{\mathrm{anonym}}$ and that of $G_{\mathrm{ref}}$, we add the following \textit{soft constraints} to the MaxSMT formulation:
\[
    \min \sum_{r} |\mathrm{degExpr}(r) - \mathrm{eDeg}(r)|
\]

Since MaxSMT-based solutions can be time-consuming and our experiments show that the greedy algorithm already achieves desirable rationality, we use the greedy approach by default in most cases.

\subsection{Configuration Generation via Mimicry}
\label{sec:design-3}
Before generating configuration files, we transform topological changes from expansion and anonymization into routing-configuration updates. For a topology from the replica method, the replication logic from the anonymization phase is used: each fake router obtains a configuration similar to its real counterpart, with matching interfaces and link settings. For topologies from the sample-connect or embedding methods, a heuristic algorithm simulating network growth configures new nodes and edges, as shown in Algorithm~\ref{alg:expand-network}.

\begin{algorithm}[!ht]
\caption{Expand Network}
\label{alg:expand-network}
\begin{algorithmic}[1]
\Require $new\_edges$
\Ensure $Nework\ Configuration$

\vspace{4pt}
\State \textbf{while} $new\_edges$ \textbf{do}
\State \quad $processed\_edges \leftarrow \emptyset$
\State \quad \textbf{for each} $(u,v)$ in $new\_edges$ \textbf{do}
\State \qquad \textbf{if} both $u$ and $v$ are new routers \textbf{then} continue
\State \qquad $processed\_edges \leftarrow processed\_edges \cup \{(u,v)\}$
\State \qquad \textbf{if} $u$ or $v$ is new router \textbf{then}
\State \qquad\quad Configure new router based on old router
\State \qquad Configure connection for $u,v$
\State \quad remove $processed\_edges$ from $new\_edges$
\State \textbf{return} $Nework\ Configuration$
\end{algorithmic}
\end{algorithm}

In each iteration, an edge \((u,v)\) from \(\textit{new\_edges}\) is processed. If both \(u\) and \(v\) are new, the edge is temporarily skipped. If only one is new, its configuration is initialized from the existing router’s settings (e.g., protocols, AS number). The link \((u,v)\) is then configured (interface assignments and link costs) and removed from \(\textit{new\_edges}\). Once all edges are handled, the updated network configuration is produced.

We now introduce the \sysname\ configuration mimicry system, which comprises two modules.

\head{Fake Configuration Generation} Existing routers have their configurations updated incrementally, preserving their original style (e.g., peer groups). For fake routers, we select a real router with highest similarity (based on degree, neighbors, AS membership, and protocols) and adopt its configuration as a template, replacing relevant fields with the fake router’s settings.

\head{Routing Policy Mimicry} For fake hosts, a similar approach is taken. Based on attributes such as egress router and AS membership, each fake host is mapped to a corresponding real host. All filter rules are examined, and if a real host matches a filter, its mapped fake hosts are included. This ensures that traffic from fake hosts is managed exactly as that for real hosts.

\subsection{SMT and Iterative Based Network Repair}
\label{sec:design-4}
We propose a network repair solution that combines SMT and iterative methods to address two issues. First, iterative methods alone can require excessive iterations and yield unstable performance in networks with added fake routers. Second, SMT struggles to precisely model complex BGP protocols and route redistribution. Our layered repair strategy first fixes routes within each AS using SMT-based constraints, and then repairs inter-AS routes by inspecting forwarding tables and adding filters directly.

\head{AS-Internal: SMT-Based Repair}
Within an AS, we encode three routing-path requirements into SMT constraints:
\begin{enumerate}
    \item For intra-AS host communications, encode the routing path between their egress routers.
    \item For a host communicating with an external AS, encode the path from the host’s egress router to the AS boundary router (ASBR).
    \item For ASBRs with interconnections, encode the routing paths among them.
\end{enumerate}
Categories (2) and (3) address complex cross-layer and cross-protocol behaviors, such as OSPF External routes relying on redistribution and iBGP routes relying on internal IP reachability. These challenges are beyond the scope of the purely iterative repair method from \confmask.

After extracting a routing path, we encode it into SMT constraints. \sysname currently supports two types of routing paths: Primary Path and ECMP Path. A Primary Path signifies that under normal conditions, all \((\text{SRC}, \text{DST})\) traffic follows a single, unique route, whereas an ECMP Path comprises multiple routes used for load balancing. Inspired by the prior CPR approach~\cite{jacobson-cpr-2017}, we define inductive SMT constraints for a Primary Path. Let \(\text{cost}^{v_1-v_2}\) denote the link cost from node \(v_1\) to node \(v_2\), \(\text{scost}^{v}\) the shortest cost from \(\text{SRC}\) to \(v\), \(\text{pred}^{v}\) the predecessor of \(v\) in the shortest path from \(\text{SRC}\), and \(\text{path}\) the required Primary Path.

\noindent (1) Positive Cost: \(\forall\, (v_1,v_2)\in E:\quad \text{cost}^{v_1-v_2} > 0 \)

\noindent (2) Source Initialization: \(\text{scost}^{\text{SRC}} = 0, \quad \text{pred}^{\text{SRC}} = \text{SRC}.\)

\noindent (3) Shortest-Path Inductive Update:
\begin{flalign*}
\text{hasAlt}^{(v_1,v_2)} :=\; &\; \exists\, v_3 \neq v_1,\; (v_3,v_2)\in E : &&\\
&\; \text{scost}^{v_3} + \text{cost}^{v_3-v_2} \;\le\; \text{scost}^{v_1} + \text{cost}^{v_1-v_2}. &&
\end{flalign*}

\noindent Then for all \(v_2 \neq \text{SRC}\) and \((v_1,v_2)\in E\),
\[
\neg \text{hasAlt}^{(v_1,v_2)}
\;\;\Rightarrow\;\;
\Bigl(
   \text{scost}^{v_2} = \text{scost}^{v_1} + \text{cost}^{v_1-v_2}
\Bigr)
\;\wedge\;
\Bigl(\text{pred}^{v_2} = v_1\Bigr),
\]

\vspace{-2em}

\[
\text{hasAlt}^{(v_1,v_2)}
\;\;\Rightarrow\;\;
\Bigl(
   \text{scost}^{v_2} < \text{scost}^{v_1} + \text{cost}^{v_1-v_2}
\Bigr)
\;\wedge\;
\Bigl(\text{pred}^{v_2} \neq v_1\Bigr).
\]

\noindent (4) Path Enforcement for the Requested Path:
\[
\forall\, i \in \{1,\dots,k\}:\quad 
\text{pred}^{\,\text{path}[i]} 
= \text{path}[i-1].
\]

\noindent For an ECMP Path, let \(\text{preds}^{v}\) be the set of all predecessors that reach \(v\) with cost equal to \(\text{scost}^{v}\), and let \(\text{paths}\) be the set of requested load-balanced routes. The inductive update for ECMP Path is defined as:
\begin{flalign*}
\forall\, v_2 \neq \text{SRC}:\;\; &
\bigvee_{v_1\in \mathrm{InEdges}(v_2)}
\Bigl(\text{scost}^{v_2} = \text{scost}^{v_1} + \text{cost}^{(v_1,v_2)}\Bigr). &&
\end{flalign*}
\vspace{-1em}
\begin{align*}
\forall\, v_2 \neq \text{SRC}:\;\;\forall\, v_1\in V:\;
& \Bigl(
    (v_1,v_2)\in E \,\wedge\, \text{scost}^{v_2} = \text{scost}^{v_1} \\ &+ \text{cost}^{(v_1,v_2)}
\Bigr) 
% &
\Longleftrightarrow\; \bigl(v_1 \in \text{preds}^{v_2}\bigr).
\end{align*}

Note that the original CPR system’s ``PC4: Primary Path'' constraint~\cite{jacobson-cpr-2017} only ensures that traffic from \(\text{SRC}\) to \(\text{DST}\) follows a specified route without distinguishing between a unique shortest path and one among several load-balanced alternatives. Using it directly in our repair process would lead to errors.
Beyond these inductive SMT encodings, \sysname leverages Counter-Example Guided Inductive Synthesis (CEGIS) from NetComplete~\cite{el-hassany-netcomplete-2018} to enhance path repair. CEGIS begins by encoding only a subset of all possible \((\text{SRC}, \text{DST})\) paths and attempts to solve them in one iteration. If the solution is invalid, the counterexample is incorporated into the constraints for the next iteration. Overall, CEGIS proves more time-efficient than purely inductive encodings in large networks.

\head{Between AS: Iterative Repair}
For inter-AS routing (e.g., BGP) crossing multiple ASes, we avoid SMT encoding of complex behaviors. Instead, we adopt the iterative repair method from \confmask, comparing forwarding tables before and after anonymization and resolving discrepancies by adding filters. Notably, if the next-hop router is an ASBR or in iBGP the local router lacks a direct Layer~3 edge\footnote{iBGP-enabled routers may rely on IGP cost to determine next-hops~\cite{cisco-bgp-2023}.}, internal network modifications are applied.

%% file: 5-implementation.tex
\begin{figure*}[t]
\begin{minipage}[t]{0.32\textwidth}
    \centering
    \includegraphics[width=\textwidth]{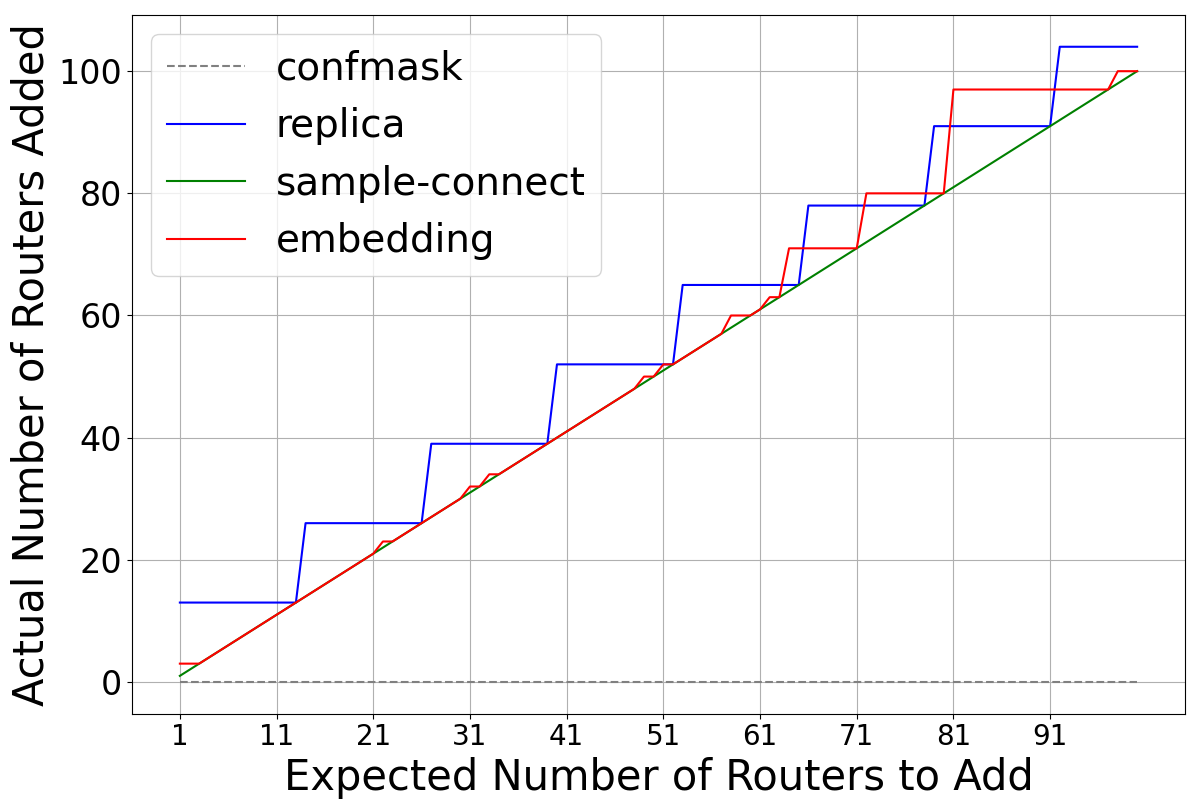}
    \vspace{-.7cm}
    \caption{\textmd{Actual number of nodes added by node addition algorithms, in network $B$}}
    \label{fig:1_node_add}
    \vspace{-1em}
\end{minipage}
\hfill
\begin{minipage}[t]{0.32\textwidth}
    \centering
    \includegraphics[width=\textwidth]{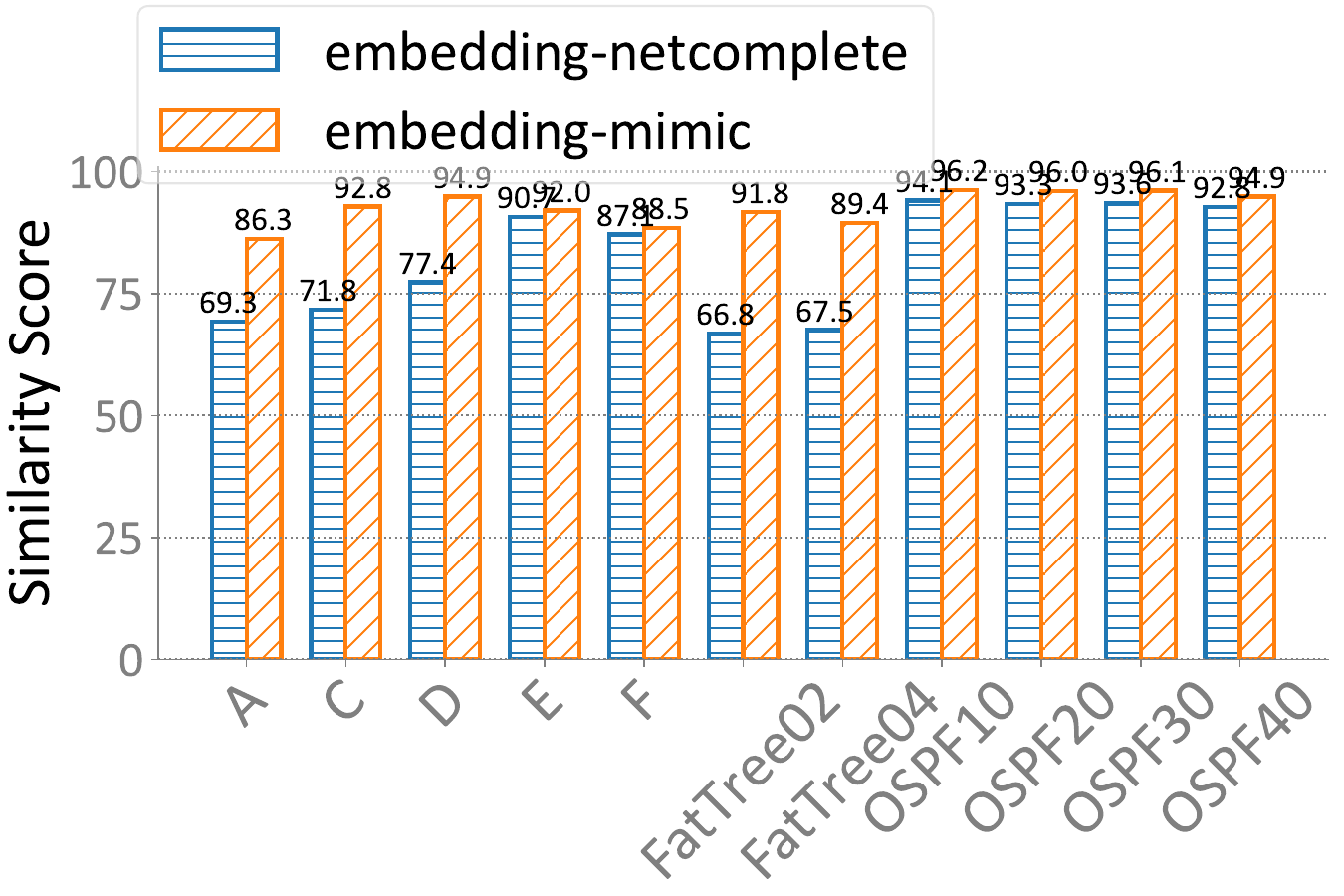}
    \vspace{-.7cm}
    \caption{\textmd{Avg. comprehensive similarity with $k_R=2$, $k_H=2$, embedding addition. \(w_{\mathrm{stanza}}\) = 0.2, \(w_{\mathrm{cmd}}\) = 0.5, \(w_{\mathrm{order}}\) = 0.3}}
    \label{fig:2_sim}
    \vspace{-1em}
\end{minipage}
\hfill
\begin{minipage}[t]{0.32\textwidth}
    \centering
    \includegraphics[width=\textwidth]{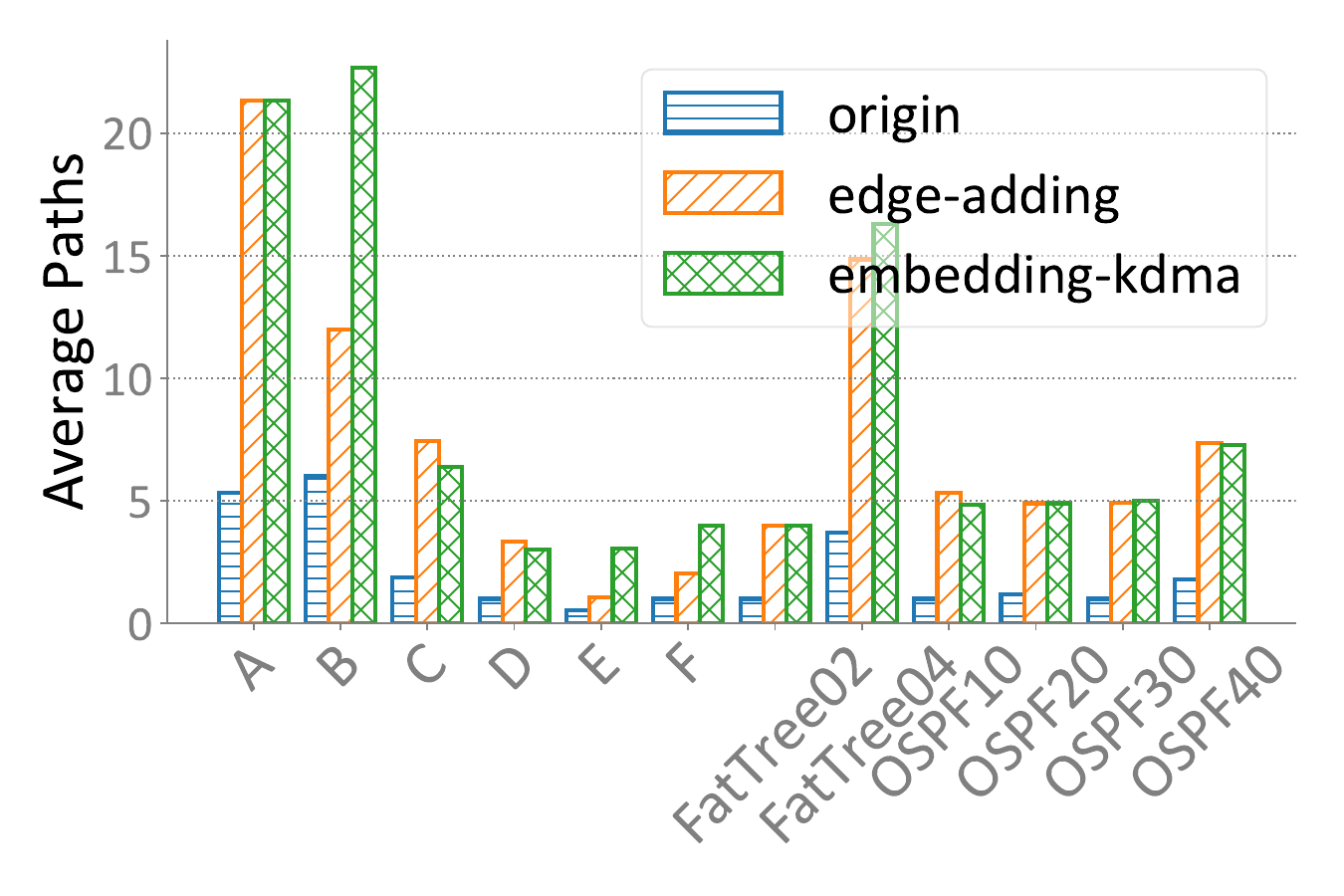}
    \vspace{-.7cm}
    \caption{\textmd{Distinct Paths among real egress routers, with $k_R=2$ and $k_H=2$}}
    \label{fig:2_nr}
    \vspace{-1em}
\end{minipage}
\end{figure*}

\section{Implementation}

We build \sysname atop \confmask and NetComplete, adding about 7k lines of Python code. Network simulation tasks such as forwarding-table computation and traceroute are handled by pybatfish, while Z3 is used for SMT constraint solving. \sysname runs end-to-end: it takes original configuration files and outputs anonymized ones. When NetComplete is needed for network synthesis, \sysname automatically provides the required inputs (network topology, routing requirements, and configuration sketches) without user intervention. Currently, \sysname supports standard Cisco commands, OSPF, BGP, and static routes; more complex features like OSPF internal-area routes remain unsupported.

\head{Route Repair}
During iterative repair, if an iBGP route uses a next hop not directly adjacent, simply adding a distribute-list at the current router has no effect. Two solutions are offered: (1) blocking the IGP route to the next hop so the router discards the iBGP route, or (2) applying a distribute-list on the next-hop router to discard the destination network route. Both are valid; (1) is faster, while (2) impacts the network less. For default routes, \sysname applies longest-prefix matching first before falling back on the default route.

\head{Configuration Mimicry}
\sysname mimics file preambles/endings, interface types, BGP address-families, and peer-group settings. It also supports standard/extended ACLs, IP-named ACLs, IP prefix lists, distribute-lists, and route-maps. 

%% file: 6-evaluation.tex
\section{Evaluation}
\label{sec:evaluation}

We conducted a series of experiments to address the following key questions:

\textbf{Question 1:} How effectively can each node-addition strategy in \sysname scale the network?\\
The replica method can only add routers in multiples of the original network size, whereas the sample-connect and embedding methods are generally more flexible in meeting users’ router-addition requirements but have upper limits on the number of routers that can be added.

\textbf{Question 2:} To what extent does the configuration-file mimicry system in \sysname improve anonymity?\\
The mimicry system produces fake configurations with higher overall similarity to the original files. By simulating routing policies, it also raises the average number of routing paths between egress routers.

\textbf{Question 3:} Is the anonymized network topology generated by \sysname a reasonable topology?\\
Using embedding-based node addition, the system yields anonymized networks with strong topological rationality, measured by the K--S distance of the degree sequence.

\textbf{Question 4:} Compared to the iterative method, what performance advantages does the SMT-based repair method provide?\\
Under randomized link costs, SMT-based repair achieves more stable and often shorter repair times than the original iterative approach.

\head{Datasets}
We tested \sysname on 12 networks covering diverse protocols and types. Networks~A--C are real-world, D is a manually configured demonstration, FatTree02--04 are classic data center topologies, and E--F plus OSPF10--40 derive from TopoZoo-based graphs synthesized via NetComplete for BGP and OSPF. 

% \vspace{-.85em}
\subsection{Ability to Expand the Network Scale}

\sysname’s methods for increasing network size sometimes cannot precisely add the user-specified number of fake routers. Fig.~\ref{fig:1_node_add} shows the gap between requested and actual routers added, with \confmask as a zero baseline. The replica method, restricted to multiples of the original router count, yields a piecewise constant function. By contrast, sample-connect and embedding are limited by the largest TopoZoo topologies; if no exact match exists, embedding may overshoot the target. As Fig.~\ref{fig:1_node_add} illustrates, embedding tracks requests closely below 60 but deviates significantly above that point. Its RMSE from the ideal curve is 4.31, compared to replica’s 7.20, indicating tighter accuracy.

Using network rescaling~\cite{mahadevan-orbis-2007} and \emph{dK}-based graph generation~\cite{mahadevan-dk-2006, gjoka-jdd-2015, gjoka-2.5k-2015} may yield larger reference topologies, potentially pushing sample-connect and embedding beyond current bounds. However, these models often assume specific properties such as power-law degree distributions, limiting broader applicability. We leave deeper investigation of these techniques for future work.

\begin{figure*}[!t]
\begin{minipage}[t]{0.32\textwidth}
    \centering
    \includegraphics[width=\textwidth]{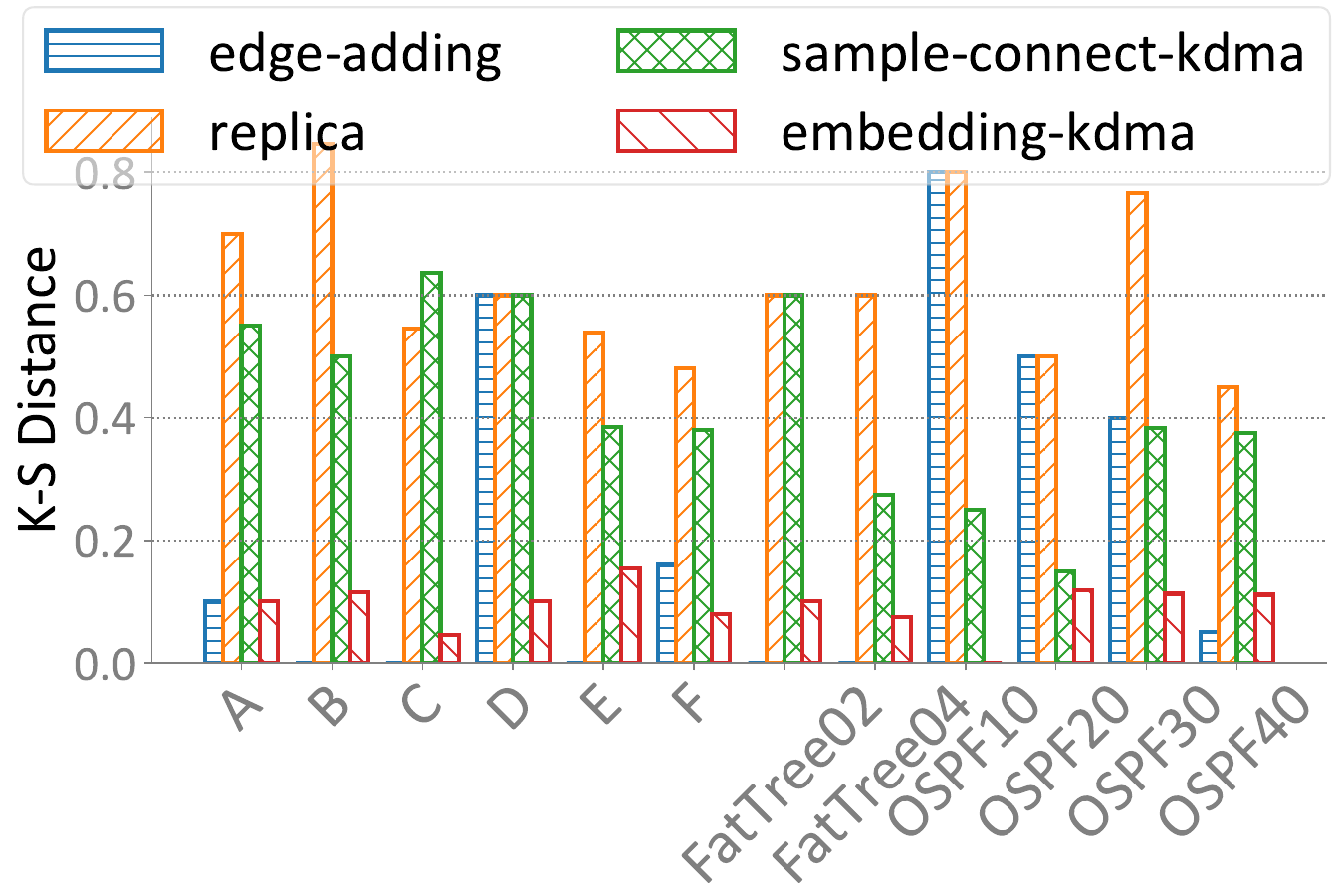}
    \vspace{-.9cm}
    \caption{\textmd{Overview of topology rationality in all networks, with $k_R=2$ and $k_H=2$}}
    \vspace{-1em}
    \label{fig:3_overview}
\end{minipage}
\hfill
\begin{minipage}[t]{0.32\textwidth}
    \centering
    \includegraphics[width=\textwidth]{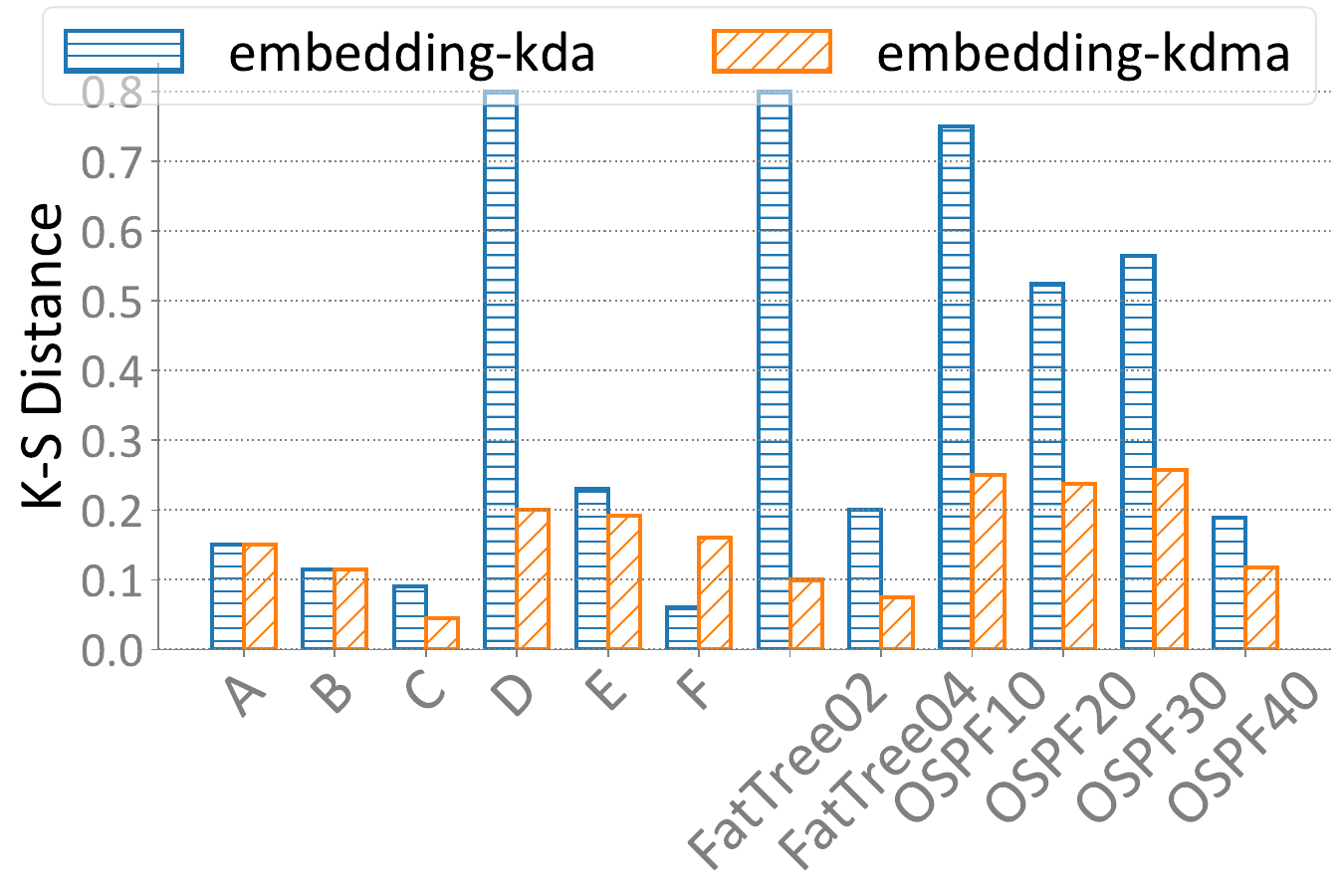}
    \vspace{-.9cm}
    \caption{\textmd{Topology rationality achieved by $k$-DMA and $k$-DA, with $k_R=4$ and $k_H=4$}}
    \vspace{-1em}
    \label{fig:3_k-DMA_k-DA}
\end{minipage}
\hfill
\begin{minipage}[t]{0.32\textwidth}
    \centering
    \includegraphics[width=\textwidth]{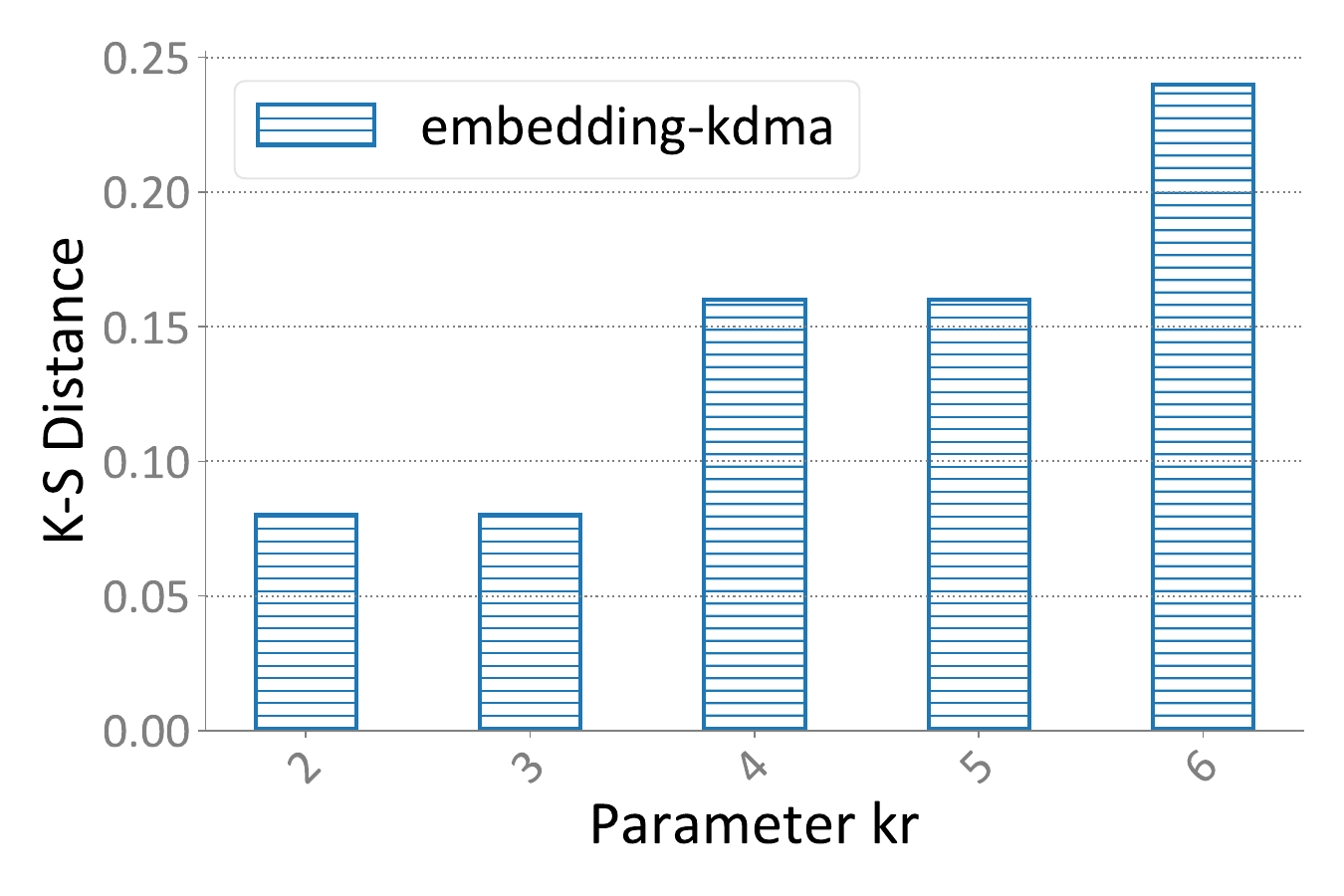}
    \vspace{-.9cm}
    \caption{\textmd{Impact of $k_R$ on topology rationality, with $k_H=2$, in network $F$, embedding addition}}
    \vspace{-1em}
    \label{fig:3_kr_F}
\end{minipage}
\end{figure*}

\subsection{Anonymity Improvement by Mimicry}

We evaluate the mimicry system’s anonymity gains using two metrics.

\head{Overall Similarity of Configurations}
A direct benefit is the ability to generate fake configurations whose style closely resembles real ones. We measure this similarity via a weighted average of three sub-metrics:
\begin{itemize}
    \item \textbf{Sim$_{\text{stanza}}$}: Cosine similarity of stanza directives, based on each directive’s proportion.
    \item \textbf{Sim$_{\text{cmd}}$}: Jaccard similarity of commands (after parameter removal) within each stanza.
    \item \textbf{Sim$_{\text{order}}$}: Similarity of stanza positional order in the file.
\end{itemize}
For each fake configuration, we compute its maximum similarity to all real configurations, then average these maxima over all fake configurations. Fig.~\ref{fig:2_sim} shows that in networks synthesized by NetComplete, the mimicry system only slightly outperforms NetComplete. However, in real-world or FatTree networks, fake configurations generated by our system achieve on average 29.03\% higher overall similarity than those produced by NetComplete, indicating closer resemblance to authentic configurations.

\head{Path Anonymity \(N_r\)}
\(N_r\) is the average number of distinct routing paths among edge routers. An indirect benefit of mimicry is \sysname’s ability to process traffic correctly in networks containing routing filters, including from fake hosts. In Fig.~\ref{fig:2_nr}, for networks (B, E, F) with filtering rules, our approach increases \(N_r\) by 96.97\% over the original \confmask on average. This substantially boosts the number of routes between real egress routers, further enhancing anonymity. 

\begin{figure*}
\begin{minipage}[t]{0.32\textwidth}
    \centering
    \includegraphics[width=\textwidth]{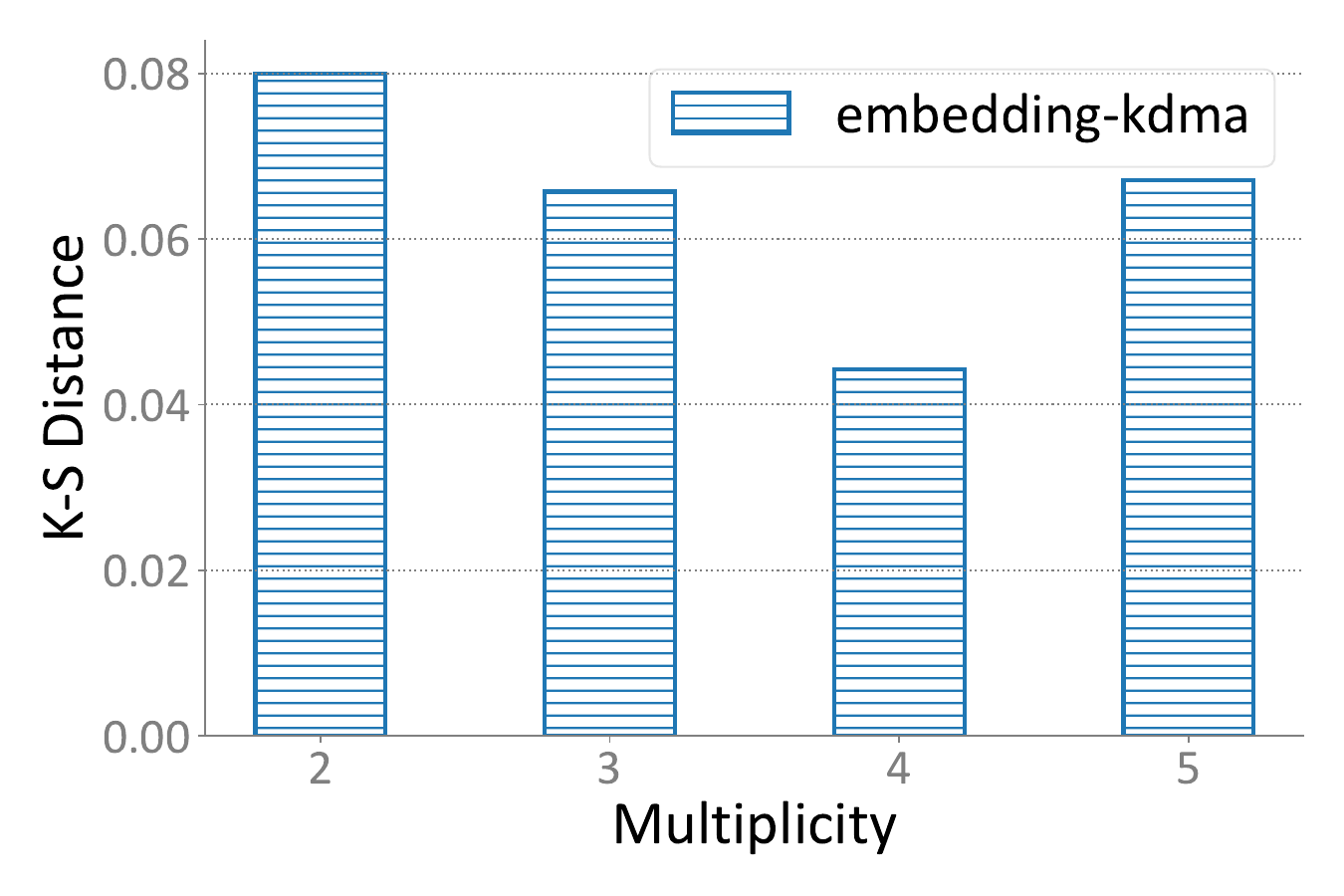}
    \vspace{-.9cm}
    \caption{\textmd{Impact of node addition multiplicity on topology rationality, with $k_R=2$ and $k_H=2$, in network $F$, embedding addition}}
    \vspace{-1em}
    \label{fig:3_mul_F}
\end{minipage}
\hfill
\begin{minipage}[t]{0.32\textwidth}
    \centering
    \includegraphics[width=\textwidth]{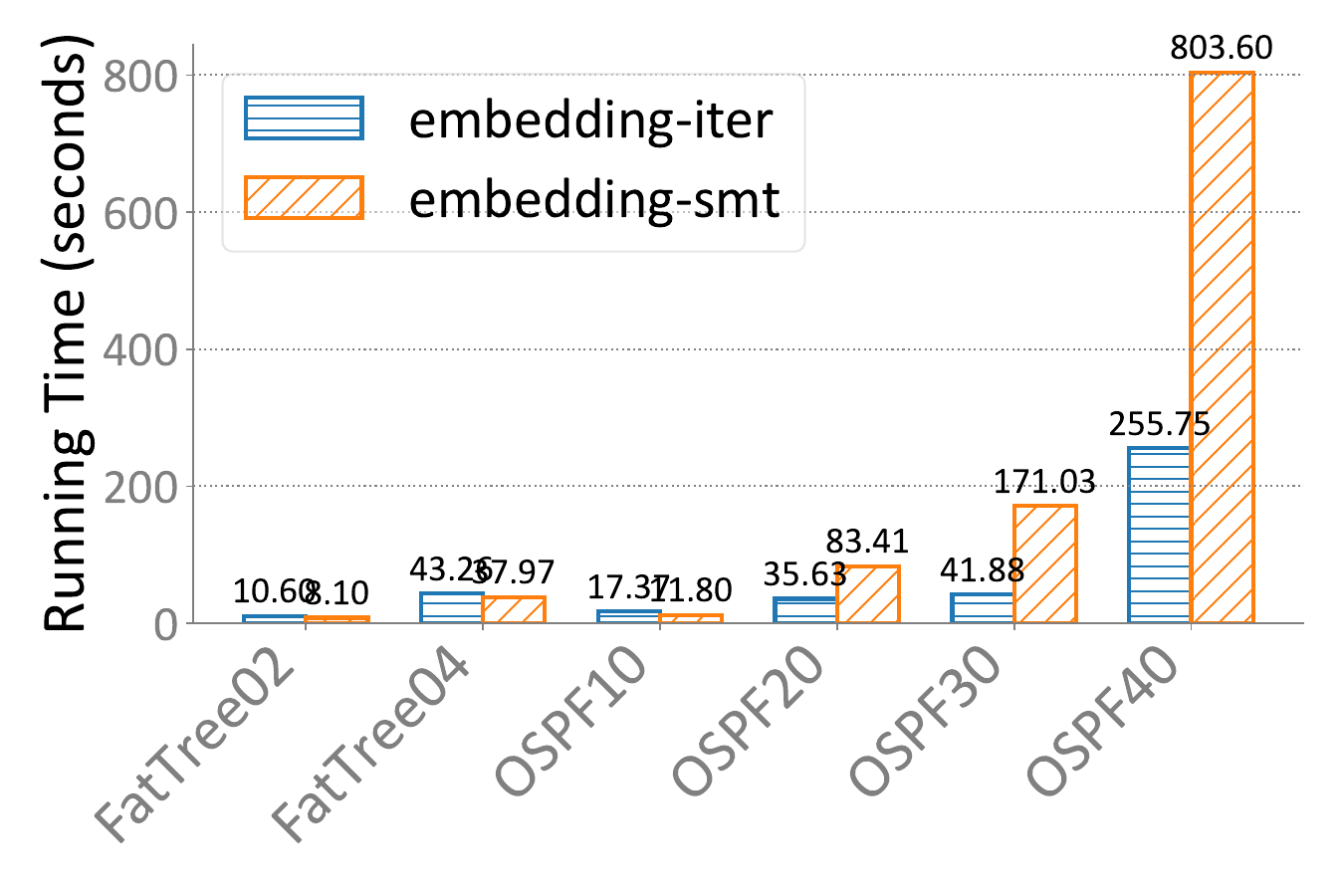}
    \vspace{-.9cm}
    \caption{\textmd{Time efficiency of SMT method and ITER method, with $k_R=2$ and $k_H=2$, default link cost, SMT method: CEGIS}}
    \vspace{-1em}
    \label{fig:4_default}
\end{minipage}
\hfill
\begin{minipage}[t]{0.32\textwidth}
    \centering
    \includegraphics[width=\textwidth]{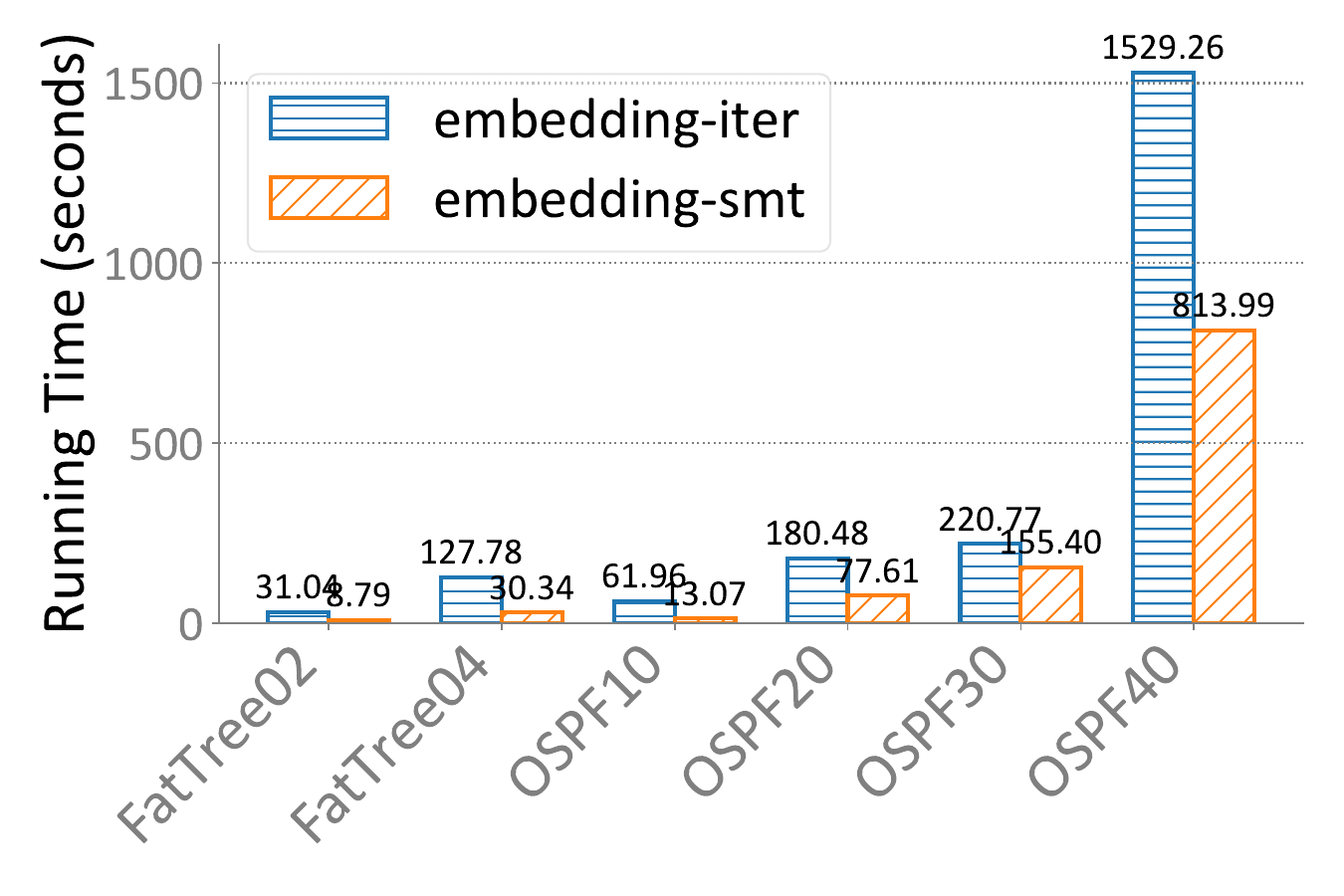}
    \vspace{-.9cm}
    \caption{\textmd{Time efficiency of SMT method and ITER method, with $k_R=2$ and $k_H=2$, random link cost(1-20), SMT method: CEGIS}}
    \vspace{-1em}
    \label{fig:4_rand}
\end{minipage}
\end{figure*}

\subsection{Network Utility Analysis}

We first evaluate the topological rationality, measured by the K–S distance of the degree sequence, for anonymized networks generated by four different methods under fixed anonymization parameters (\(k_R=2\), \(k_H=2\), and \(\text{mul}=2\)). For the edge-adding and replica methods, rationality is measured relative to the original network, whereas for the sample-connect and embedding methods, it is measured relative to the reference topology. As shown in Figure~\ref{fig:3_overview}, \sysname’s edge-adding method exhibits extreme behavior: it produces a K–S distance of 0 in FatTree networks (indicating an unchanged topology and hence no anonymization) while reaching values up to 0.8 in smaller networks (e.g., OSPF10), reflecting low topological rationality. In contrast, the replica method consistently yields K–S distances exceeding 0.4 and, in some cases, above 0.8, again indicating poor rationality. Both sample-connect and embedding methods attempt to integrate the original network with a reference topology; however, because the embedding method explicitly minimizes differences in the degree sequence during graph construction, it consistently achieves higher topological rationality with K–S distances below 0.2, significantly outperforming the random linkage strategy of sample-connect.

Next, we compare the topological rationality of networks produced by the new $k$-DMA anonymization algorithm with those generated by the $k$-DA algorithm within the embedding framework. As illustrated in Figure~\ref{fig:3_k-DMA_k-DA}, $k$-DMA generally outperforms $k$-DA in most networks—especially in smaller ones (e.g., D, FatTree02, OSPF10)—where $k$-DA produces larger disparities in the degree sequence. Overall, $k$-DMA proves to be a practical solution when attackers have limited prior knowledge and a high level of topological rationality is required.

Finally, we investigate how topological rationality under the $k$-DMA algorithm varies with the anonymization parameter \(k_R\) and the number of added nodes. Figure~\ref{fig:3_kr_F} shows that the K–S distance increases with \(k_R\) (with a correlation coefficient \(r = 0.94\)), indicating a strong positive correlation. However, Figure~\ref{fig:3_mul_F} reveals only a weak negative correlation (\(r = -0.52\)) with the number of added nodes. This variability arises because changes in the number of added nodes lead to adjustments in the reference topology, complicating direct comparisons of the embedding process across different expansion scales.

\subsection{Time Efficiency of SMT-based Methods}

When all OSPF link costs are set to 1 or use the default (1), results in Fig.~\ref{fig:4_default} show the SMT-based repair method is less efficient for medium and large networks compared to the iterative approach. This aligns with prior work~\cite{jacobson-cpr-2017} indicating that, without parallelization, SMT scales poorly for larger topologies.\footnote{Because link-cost encoding is required for SMT, flow-based parallel optimizations cannot apply~\cite{jacobson-cpr-2017}.}

However, under randomized OSPF costs between 1 and 20 (Fig.~\ref{fig:4_rand}), the comparative efficiency changes. The iterative method demands more iterations and incurs notably higher repair time, whereas the SMT-based approach remains largely unaffected, sustaining relatively stable performance. In these scenarios, SMT-based repair can even surpass the iterative method. Overall, unless one must repair routing in an extremely large network, the SMT-based strategy provides more stable and predictable performance. By contrast, the iterative method is more sensitive to fluctuations from specific link-cost configurations.

%% file: 7-related.tex
\section{Related Work}

\head{Configuration Anonymization and Sharing}
Traditional configuration file anonymization systems focus on obfuscating sensitive fields like IP addresses, AS numbers, and management information~\cite{maltz-configanonym-2004, intentionet_netconan_2023, han-confighub-2024,xu_prefix-preserving_2002}. \confmask is the first system to anonymize network topologies while preserving functional equivalence~\cite{wang-confmask-2024}, but it does not support scaling the network. ConfigHub provides a Data Expansion module that generates larger configuration files by randomly linking topologies and discarding original routing specifications via NetComplete. This random expansion results in lower fidelity and practical utility since the expanded networks no longer reflect real routing characteristics.

\head{Graph Anonymization}
Graph anonymization aims to achieve \emph{k}-anonymity using methods such as \emph{k}-degree anonymity~\cite{liu-kDA-2008,lu-fast-kDA-2012}, \emph{k}-neighborhood anonymity~\cite{zhou-kNA-2008}, \emph{k}-automorphism~\cite{zou-kauto-2009}, and \emph{k}-isomorphism~\cite{cheng-kiso-2010, takbiri-replica-2019}. These techniques generally add or edit edges to mask sensitive structures. However, when applied to topologies expanded with extra nodes, they may cause excessive modifications. Differential privacy~\cite{dwork-dp-2006} has also been explored to anonymize node~\cite{day-dp-node-2016, jian-dp-node-2023} and edge~\cite{sala-dp-pygmalion-2011, jorgensen-dp-edge-2016} information.

\head{Network Configuration Synthesis}
Several systems synthesize network configurations from high-level requirements, including SyNET~\cite{el-hassany-syNET-2017}, Propane~\cite{Beckett-propane-2016}, NetComplete~\cite{el-hassany-netcomplete-2018}, and CPR~\cite{jacobson-cpr-2017}. SyNET uses stratified Datalog but offers limited BGP support. Propane targets BGP with a high-level language for policy enforcement, while NetComplete and CPR use SMT-based methods; CPR employs an abstract control plane (ARC)~\cite{jacobson-arc-2016} to support route redistribution, though its BGP modeling remains constrained. These synthesis systems aim to automate configuration generation and reduce manual errors, yet they often face trade-offs between expressiveness and computational efficiency.

\head{Network Verification}
Systems such as~\cite{fogel-batfish-2015,jacobson-arc-2016,brikner-config2spec-2020,abhashkumar-tiramisu-2020,beckett-minesweeper-2017,ye-hoyan-2020} facilitate control-plane simulation, verification, and configuration error repair. The anonymized configurations produced by \sysname are compatible with these tools, enabling verification while preserving confidentiality. Such verification is critical for ensuring that network configurations perform as intended.

% \vspace{-4em}

%% file: 8-conclusions.tex
\section{Discussion}
\label{sec:discussion}

\head{Different anonymity definition}
\(\epsilon\)-differential privacy is promising in graph anonymization. However, applying existing node- and edge-based \(\epsilon\)-DP approaches directly is challenging because our anonymized graph must retain the original nodes and edges, requiring knowledge of the raw topology during post-processing. This conflicts with the \(\epsilon\)-DP post-processing immunity theorem, valid only for data-independent transformations. Future studies could devise new formulations of graph \(\epsilon\)-DP tailored to preserve structural features while ensuring robust anonymity guarantees for real-world network data.

\head{Topology rationality definition based on dK-series}
The dK-series framework describes subgraph degree correlations at multiple dimensions: 0K for average degree, 1K for degree distribution, 2K for joint degree distribution, and 3K for triangles or wedges. \sysname currently employs the 1K property for topology rationality. Incorporating higher-dimensional dK-series would reveal deeper structural patterns, but necessitates updating our graph embedding and \emph{k}-DMA techniques. By capturing more granular dK properties, \sysname could further improve anonymization quality while preserving realistic topological features.

\head{Cross-layer protocols and Traffic anonymization}
Currently, \sysname anonymizes primarily at the network layer and control plane. In practice, adversaries may exploit traffic patterns or higher-layer metadata to correlate and de-anonymize real topologies. Future work could integrate cross-layer anonymization—e.g., Layer~2 VLAN settings or protocols like HTTP and TLS—with traffic anonymization. Such multi-level strategies can offer more robust protection against sophisticated attacks, ensuring that both routing and traffic information remain hidden from adversarial inference.

\vspace{-.6em}
\section{Conclusion}
\label{sec:conclusion}

Modern anonymization approaches for network configuration files often fail to effectively obscure network scale. To address this gap, we present \sysname, a system that adaptively adds nodes and links on top of the original topology, generates router configurations matching real styles, and employs a layered routing repair framework to preserve functional equivalence. By combining a graph-embedding-based topology expansion with a new \emph{k}-degree mapping anonymity definition, \sysname effectively masks node-degree information while minimizing structural distortions. Experimental results on real and emulated networks demonstrate \sysname's advantages in topological rationality, configuration fidelity, and repair efficiency, thereby enabling safer and more practical configuration sharing.